\documentclass[preprint,12pt]{elsarticle}

\usepackage{amssymb}
\usepackage{lineno}
\usepackage[normalem]{ulem}
\usepackage{color}
\journal{Astroparticle Physics}

\begin{document}

\begin{frontmatter}



\title{The HiSCORE concept for gamma-ray and cosmic-ray astrophysics beyond 10\,TeV}


\author[1]{Martin Tluczykont\corref{cor1}}
\cortext[cor1]{tel.: +49+40\,8998\,2993, fax: +49+40\,8998\,2170}
\ead{martin.tluczykont@physik.uni-hamburg.de}
\author[1]{Daniel Hampf}
\author[1]{Dieter Horns}
\author[1]{Dominik Spitschan}
\author[2]{Leonid Kuzmichev}
\author[2]{Vasily Prosin}
\author[3]{Christian Spiering}
\author[3]{Ralf Wischnewski}
\address[1]{Institute for Experimental Physics, University of Hamburg, Luruper Chaussee 149, 22761 Hamburg}
\address[2]{Skobeltsyn Institute for Nuclear Physics, Lomonosow Moscow State University, 1 Leninskie gory, 119991 Moscow}
\address[3]{DESY, Platanenallee 6, 15378 Zeuthen, Germany}

\begin{abstract}
   Air-shower measurements in the {primary} energy range beyond 10\,TeV 
   can be used to address {important} questions of 
   astroparticle and particle physics.
   The most prominent among these questions are the search for the origin of charged Galactic
   cosmic rays and {the so-far little understood transition
   from Galactic to extra-galactic cosmic rays. A very promising avenue towards answering these fundamental questions
   is the construction of an air-shower detector with sufficient sensitivity for gamma-rays to {identify} the accelerators
   and large exposure to achieve accurate spectroscopy of local cosmic rays.} 
   With the new ground-based large-area (up to 100\,km$^2$) wide-angle ({$\Omega\,\sim$0.6--0.85\,sr})
   air-shower detector concept HiSCORE (Hundred*i Square-km Cosmic ORigin Explorer), we aim at exploring
   the cosmic ray and gamma-ray sky (accelerator-sky) in the energy range from
   {few 10s of TeV} to 1\,EeV {using the non-imaging air-Cherenkov detection {technique}}. 
%   This energy regime is crucial for a solution of the origin of cosmic rays and also
%   allows to address several other fundamental particle and astro-particle physics questions.
%   Currently existing and up-coming instruments only cover (or will cover) the gamma-ray observation window
%   up to few 100\,TeV. Going further up in energy requires significantly larger instrumented areas
%   than previously achieved or planned.
   The full detector simulation is presented here.
   The resulting sensitivity of a HiSCORE-type detector to gamma-rays
   will extend the energy range so far accessed by other experiments beyond energies of 50\,--100\,TeV,
   thereby opening up the ultra high energy gamma-ray (UHE gamma-rays, E$>$10\,TeV) observation window.
\end{abstract}


\end{frontmatter}

%\linenumbers

\section{Introduction}
\label{hiscore_introduction}
% main question of astroparticle physics: origin of cosmic rays
% ... gamma-rays: only up to 100TeV
% ... cosmic rays: need more statistics
%
%%% we have developed a concept for a new wide-angle large-area cosmic ray and gamma-ray detector: HiSCORE
The current knowledge about the origin of cosmic rays has been accumulated
following two different approaches: 
(i) by measuring  in detail the energy spectrum and mass composition of 
the local cosmic-ray population  and (ii) by gamma-ray ($E>100$~MeV)
observations of both individual astrophysical objects as well as
the diffuse emission from the interstellar medium. Both approaches provide complementary information/constraints
on the most relevant quantities:
e.g. 
the measurement of spallation products and cosmo-genic nuclei provides
information on the energy dependence of cosmic-ray transport and 
the escape time of cosmic rays out of the Galaxy. 
Gamma-ray observations constrain the spatial distribution and 
properties of the cosmic ray
accelerators and  the density of cosmic rays in the interstellar medium.  

Cosmic-ray measurements
through air-shower techniques are the only means to collect 
sufficient event statistics to measure cosmic rays at energies close to the
knee ($\approx 3\times 10^{15}~$eV) 
in the all-particle energy spectrum. The traditional air-shower detectors
sample the lateral density function (LDF) of secondary particles {or photons} on the ground.
Given the large intrinsic fluctuations in the shower development
and
{that only a small fraction of the particles are sampled}
%the sparse sampling
($\approx 10^{-4}$), the energy resolution 
and sensitivity to different primary particles is rather limited. 
Combining detection of different components of the air shower as e.g.
realised in the KASCADE air shower field \cite{2004ApJ...608..865A}, improves the situation considerably 
but suffers from limited collection area. Established techniques to follow the
longitudinal air shower development include muon tracking, air Cherenkov, 
and air fluorescence observations.

The latter technique has been realised 
quite early and remains one of the most sensitive techniques at ultra-high
energies \cite[Linsley, Fly's eye, HiRes, Pierre-Auger Observatory,
Telescope-Array, see][and references therein]{2011ARA&A..49..119K}. 
The non-imaging air Cherenkov technique measures the arrival time and the LDF of the Cherenkov
photons in the air shower front. This technique
%%%%%%%%%%%%%%%%%%%%%REMOVED:%%%%%%
% has
%%%%%%%%%%%%%%%%%%%%%%%%%%%%%%%%%%%  
is sensitive to the longitudinal air shower development (mainly position of the shower maximum)
as demonstrated with e.g.
Themistocle \cite{1993APh.....1..341T},
AIROBICC \cite{1995APh.....3..321K},
%Blanca \cite{1997ICRC....5..189C,2001APh....15...49F}, 
Blanca \cite{2001APh....15...49F}, 
%Tunka \cite{2005ICRC....8..255B},
Tunka \cite{2012NIMPA.692...98B},
Jakutsk \cite{1986NIMPA.248..224D}. 
The longitudinal air shower development  is sensitive to the initial particle species. 
Both techniques allow a comparably good energy resolution which suffers
less from the fluctuations and the limited sampling.  A number of new approaches
for air shower detection have been proposed and partially tested including long-wavelength {(MHz)} radio
measurements\footnote{the dominant emission processes are geo-synchrotron and charge separation} \cite[see e.g.][]{2012NIMPA.662S..72H}, and 
molecular Bremsstrahlung emission at {GHz frequencies}
%%%%%%% REMOVED %%%%%%%%%%%%%
% micro-meter wavelengths
%%%%%%%%%%%%%%%%%%%%%%%%%%%%%
\cite{2008PhRvD..78c2007G,2013EPJWC..5308010S}.
In the sense of shower-front sampling, the long-wavelength
radio observations are comparable to the air Cherenkov technique  while the
molecular Bremsstrahlung has {analogies} to the air fluorescence (mostly
isotropic emission) and {would allow} for imaging of the air shower development.
%Both
%techniques have been demonstrated successfully at energies of $10^{17}$~eV and
%above

For approach (i) -- spectroscopy
%(good energy resolution)
and measurement of chemical composition 
of cosmic rays in the energy range from below the
knee to the ankle ($10^{18}$~eV) -- the air Cherenkov approach appears to be {among} the best choice{{s}},
{considering its good energy resolution of the order of 10\,\%, and a typical resolution of the shower maximum
of the order of 30\,g/cm$^2$ \cite{2012NIMPA.692...98B}.
See also Section~\ref{section:air_shower_reconstruction}}.
For approach (ii) --  gamma-ray 
observations -- the currently most succesfull technique is the imaging air Cherenkov technique with multiple telescopes 
(imaging air Cherenkov telescopes: IACTs).
A large array of IACTs is currently under design to achieve a ten-fold improvement in flux sensitivity
{as compared to current generation instruments}: the Cherenkov
Telescope Array \cite[CTA, see][]{2011ExA....32..193A}.
Nevertheless CTA is designed to achieve optimum sensitivity at TeV energies and
will suffer from its limited collection area at energies
%%%%%%%%%% REMOVED %%%%%%%%%%%%%
%well beyond ten TeV.
%%%%%%%%%%%%%%%%%%%%%%%%%%%%%%%%
{beyond 100\,TeV.}
The non-imaging air Cherenkov technique allows to extend the collection area to {several} square kilometers with a moderate
number of read-out channels\footnote{there are complementary approaches using large
field of view cameras which would allow to increase the spacing of individual telescopes \cite{2008NIMPA.588...48R}}.
In combination with the demonstrated good angular resolution in non-imaging Cherenkov air shower arrays,
%THEMISTOCLE:1991AIPC..220..237B} 
%and AIROBICC \cite{1995APh.....3..321K})
a multi-km$^2$ array with good sensitivity above 10 TeV appears feasible and is explored here.

{
With the Hundred*i Square-km Cosmic ORigin Explorer HiSCORE, we want to cover both approaches (i) and (ii) described above.
A central question will be the search for the elusive
pevatrons \cite{2007ApJ...665L.131G}, the accelerators of cosmic rays up to the PeV energy regime.
}
For more details on physics topics for HiSCORE, see \cite[][]{2011AdSpR..48.1935T} and references therein.

\section{HiSCORE detector design}
\label{hiscore_detector_design}
\subsection{Detector array layout}
HiSCORE will consist of an array of wide-angle light-sensitive detector
stations, distributed over an area of the order of 100\,km$^2$.
As compared to previous experiments, important aspects of HiSCORE are
different (see Table~\ref{score_paper_properties}): 
an instrumented area larger by more than an order of magnitude,
up to a factor 16 larger light-collecting area per station, and the usage of fast GHz {waveform sampling} electronics.
\begin{table*}[th]
\caption{\label{score_paper_properties}Basic design characteristics of the HiSCORE detector in
comparison with other experiments.
The total instrumented area $A$, the light collection area $a$ of an individual station,
 the field of view $FoV$, the inter-station distance $d$ and the number of detector stations $N$
are listed.  }
\centering
\begin{tabular}{llllll}\hline

Parameter:	& $A$ 		& $a$ 		& $FoV$ 	& $d$		& $N$	\\
Unit		& $[\mathrm{km}^2]$ 	& $[\mathrm{m}^2]$	& $[\mathrm{sr}]$	& $[\mathrm{m}]$		& 	\\
\bf HiSCORE	& \bf 100     	& \bf 0.5     	& \bf 0.60--0.85    	& \bf 150$^a$	& \bf 4489	\\
Tunka-133     	& 1$^{b}$	& 0.031		& 1.8           & 85            & 133		\\
 Blanca        	& 0.2   	& 0.1   	& 0.12  	& 35    	& 144   	\\
AIROBICC 	& 0.04     	& 0.13     	& 1        	& 15--30       	& 49       \\
 Themistocle 	& 0.08		& 0.5		& 0.008		& 50--100	& 18	\\\hline
\end{tabular}
\flushleft
{\footnotesize $^a$ Inter-station spacing used for the simulation results presented in the present paper
are not optimized yet.}\\
{\footnotesize $^{b}$ In 2011, the effective area for high energy events was increased to 3\,km$^2$ by extending 
the array with additional 42 optical detectors, placed at a distance of 1\,km from the array center \cite{2012NIMPA.692...98B}.}
\end{table*}

\begin{figure}[!t]
  \centering
  \includegraphics[width=\columnwidth]{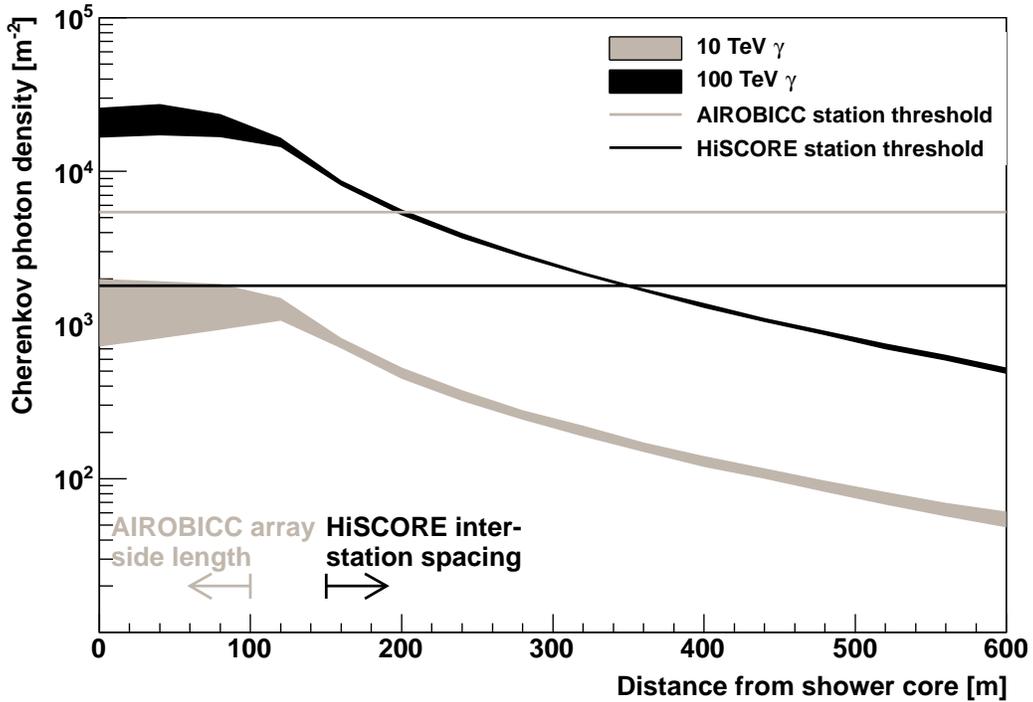}
  \caption{Lateral photon density function (LDF) of Cherenkov light for airshowers at sea level
           initiated by a 10\,TeV gamma-ray shower (grey area) and a 100\,TeV gamma-ray shower.
           The {light} sensitivity level of one HiSCORE detector station is indicated by the solid line.
           For comparison, the corresponding {light} sensitivity level for AIROBICC is also shown (grey line).
	Note, the observation level of AIROBICC was 2\,200\,m with a correspondingly higher photon density.}
  \label{score_paper_lateral}
 \end{figure}

Since we aim at a very large instrumented area, a low array density with large inter-station spacings is favoured.
Figure~\ref{score_paper_lateral} shows the lateral photon density function (LDF) of Cherenkov light on
the ground.
Within a radius of 120\,m around the shower core position,
the LDF is roughly constant.
%\sout{but shows a large spread from shower to shower. Fluctuations are much lower beyond 120\,m.}
{Beyond 120\,m, the photon density decreases following a power law.}
With a station spacing of 100\,m or more (depending on the array layout and partly variable),
HiSCORE will primarily measure the outer part of the LDF,
i.e. most stations will sample the LDF beyond 120\,m distance from the shower core.
Still, a few stations will lie within the central 120\,m of the lightpool.
Due to the low Cherenkov photon density far away from the shower core, a large light collection area
$a$ of the individual detector stations is needed.
{With {the standard array configuration and} a chosen area $a = 0.5$\,m$^2$,
{on average 3} stations within 120\,m of the shower core
are {found} to be above threshold for showers {at 50\,TeV} primary energy.
{The energy threshold for gamma-rays at reconstruction level therefore is 50\,TeV.
When using alternative layouts with partly higher station densities or larger PMTs
(see section~\ref{alternative_configurations}), the energy threshold at reconstruction
level can be reduced towards our aim of 10\,TeV.}
At 100\,TeV, the LDF can be sampled in the power law part up to core distances of 450\,m.}
In Figure~\ref{score_paper_lateral}, the basic difference in scale becomes apparent: The inter-station spacing of
the HiSCORE array is of the same order of magnitude as the total side-length of the AIROBICC detector.
%
%%%%%%%%%%%%%%%%%%%%%%%
%
\subsection{Detector station}
\begin{figure}[ht!]
 \centering
 \includegraphics[width=\columnwidth]{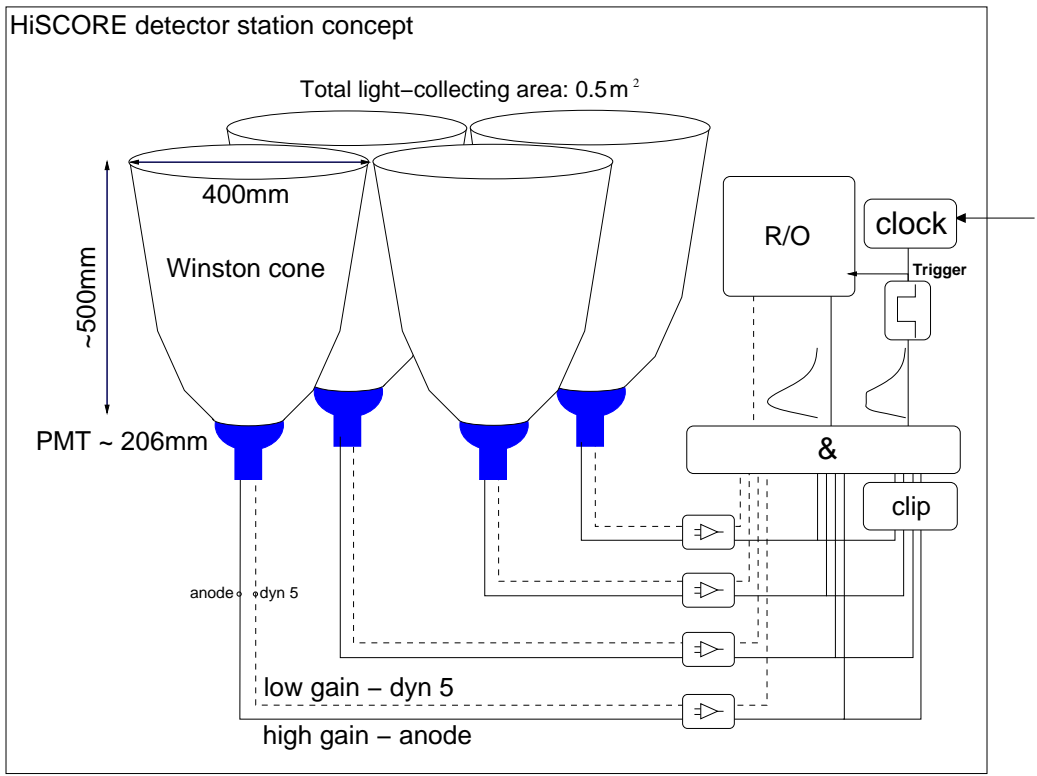}
 \caption{%\sout{Left: }
	Schematical layout of a HiSCORE detector station: Four PMTs each equipped with
         a Winston cone and read-out electronics are encased in an aluminum box with a sliding lid (not shown).
To increase the dynamic range, the PMT anode and the next to last dynode are read out.
%An industrial mini computer inside the station is connected to the data acquisition center and the slow control
%system. A precision clock, synchronized to the central array clock, assures (sub-) ns relative inter-station timing over the whole HiSCORE array.
         }
 \label{score_paper_detector_station_concept}
\end{figure}

A HiSCORE detector station consists of four photomultiplier tubes (PMTs),
each equipped with a light-collecting Winston cone of $\approx$30$^\circ$ half-opening angle pointing to the zenith.
%All four modules (PMT+cone), including the trigger, read-out electronics, and communication
%are integrated in a station box equipped with a sliding lid.
The advantages of using four PMT channels per station are on the one hand 
the suppression  of random triggers from nightsky background (NSB) light through a local coincidence trigger
condition, and on the other hand an increase of the light collecting station area ($a$).
When using four 8'' PMTs and a height of 0.5\,m of the Winston cones $a~=~0.5\,\mathrm{m}^2$
is achieved.
A schematical view of the station concept is given in Figure~\ref{score_paper_detector_station_concept}.
%

% gain, dynamic range
The PMTs have to fulfill two basic requirements: The operational gain must be such that
the anode current stays within manufacturer limits under the expected NSB conditions.
With a nominal gain of 10$^4$, the 6-stage PMT KB9352 from Electron Tubes fulfills this requirement.
%\sout{Alternatively,}
A modified R~5912  Hamamatsu PMT with six dynodes 
{is} {an} interesting {alternative}.
The dynamic range has to be as high as 10$^5$, since we aim at measurements between 10\,TeV and 1\,EeV.
This {could} be achieved by reading out one or two dynodes in addition to the anode signal.
{
Assuming a voltage range of the readout of 14\,mV to 1\,V (e.g. DRS4 evaluation board, \cite{2010NIMPA.623..486R}),
the anode (high gain, first readout stage) channel provides a dynamic range of 70.
At the upper end of this voltage range, the actual anode signal is 100\,mV
(when using a premplification of a factor 10), i.e. well within the linear anode voltage range
which extends up to roughly 2\,V. With a dynode (low gain, second readout stage) channel at a
factor 50 lower amplification, sufficient overlap between both channels is given.
The total dynamic range then is 3500.
% (for a theoretical anode signal ranging from 14\,mV to 50\,V).
An additional, second dynode readout would provide a further extension of the dynamic range up to 10$^5$.
Alternatively, without a third readout stage, events at the highest energies could also be reconstructed
using stations far away from the shower core, and applying appropriate low weights for the inner (saturated) stations.
}
{While such methods of increasing the dynamic range are feasible, they will require careful calibration and will ultimately be
a source of systematic errors towards the high energy end of the sensitive detector range.}

The Winston cones can be conveniently built from segments of reflective
%\sout{foil}
{aluminum layer} on a synthetic, flexible carrier material (Alanod 4300UP).
Ten segments are cut out from {this} material
{and} assembled similarly to a barrel.
The Winston cone shape is well reproduced along the optical axis.

The inside of the station box is equipped with slow-control electronics, a lid motor, 
high-voltage (HV) systems, a local station trigger,
and a read-out system (pre-amplifiers, analog signal sampling board).
A fast signal read-out and digitization in the GHz regime is provided
by {a read-out board with up to 8 channels based on the DRS\,4 chip \cite{2010NIMPA.623..486R},}
%by a DRS\,4 chip
with a depth of 1024 cells at a resolution of up to 0.2\,ns per cell (5\,GHz). In order to retain a sufficiently
wide read-out window, a sampling frequency of 1\,GHz is used corresponding to a read-out window of 1\,$\mu$s.

For a large array such as HiSCORE, with large inter-station spacings, accurate relative timing is
crucial.
Simulations show that a relative timing accuracy {between stations} of the order of 1\,ns is required to optimize
the angular resolution as well as the reconstruction accuracy of the position of shower maximum. 
Systems for relative synchronisation at the (sub)-ns accuracy level are currently under study. Possible
approaches are
{WhiteRabbit (PTP over synchronuous ethernet \cite[][]{white_rabbit}), with sub-ns resolution reached in the HiSCORE prototype \cite[][]{brueckner_wr,brueckner_wr_2,wischnewski_wr}},
%\footnote{{\tt http://www.sevensols.com/whiterabbitsolution/}}.
or a system based on the ethernet carrier frequency such as used by the Tunka-133 array \cite{2008apsp.conf..287B}, 
or radio signal phase synchronisation \cite{2010NIMPA.615..277S}.

% trigger
\subsection{Trigger and read-out}
\begin{figure}[ht!]
 \centering
 \includegraphics[width=\columnwidth]{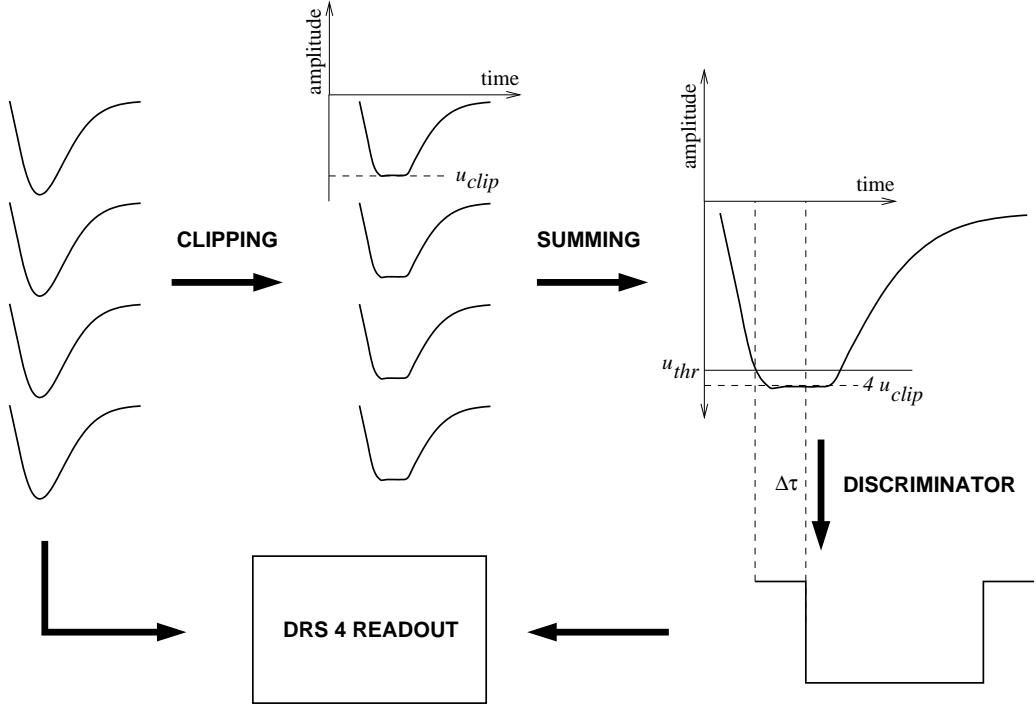}
 \caption{
		Trigger scheme for the local station: Analog signals are clipped to
		a level $u_\mathrm{clip}$ before summing. 
		The sum of clipped signals passes a discriminator with
		threshold $0<u_\mathrm{thr}/u_\mathrm{clip}<4$, 
		which finally triggers the DRS\,4 read-out.
         }
 \label{trigger_scheme}
\end{figure}

\begin{figure}[!ht]
  \centering
  \includegraphics[width=\columnwidth]{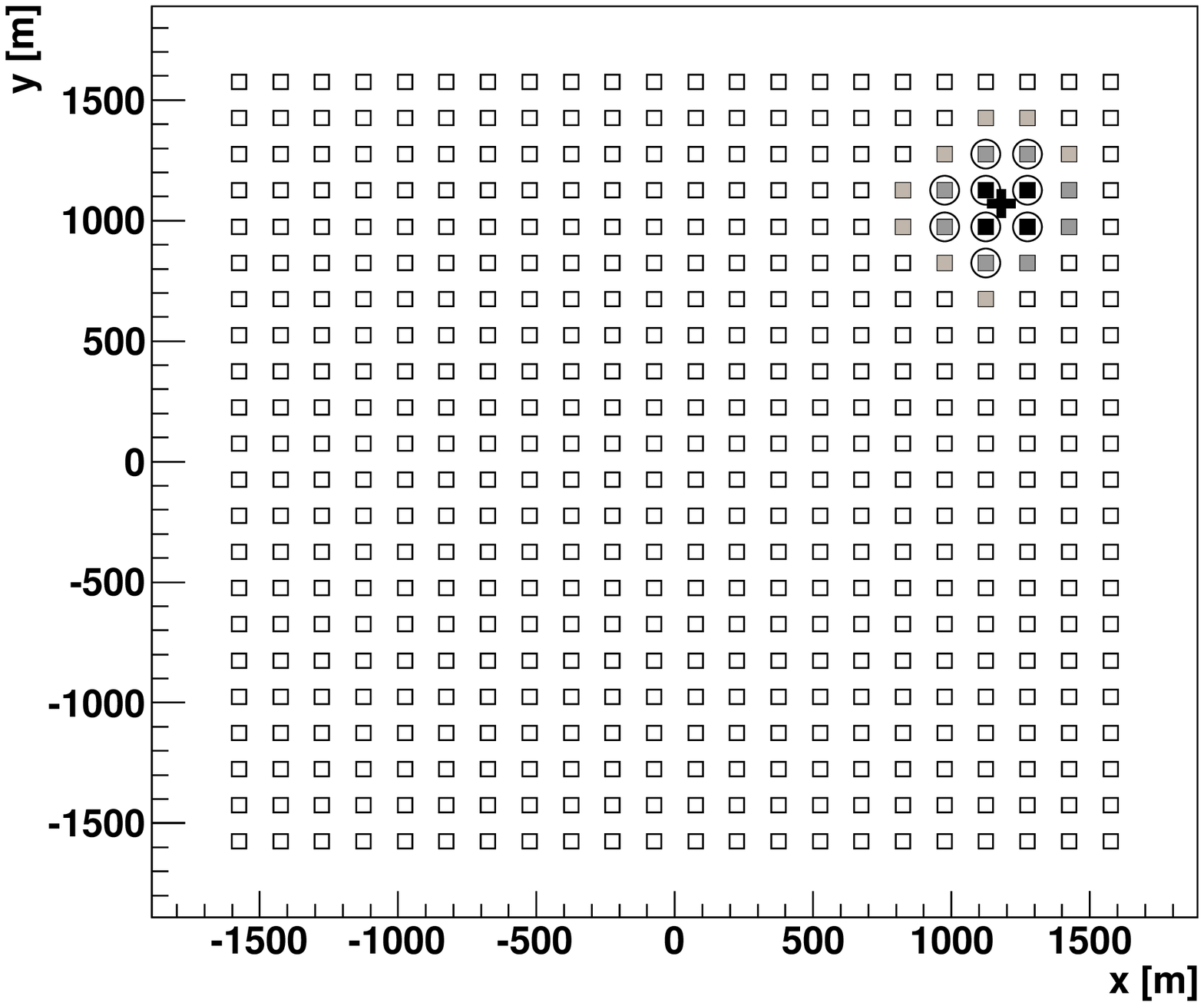}
  \caption{HiSCORE event display of a 187\,TeV gamma-ray event recorded by the standard geometry array configuration.
{Recorded photons in the detector stations are binned in high (black), medium (dark grey) and low (ligh grey) densities.}
Triggered stations are marked by an additional
%\sout{grey}
circle.
{The cross indicates the simulated position of the shower core.}
  \label{event_display}}
 \end{figure}

%\sout{The envisaged trigger scheme encompasses up to two levels.}
%
The
%\sout{first trigger level}
{station trigger} is illustrated in Figure~\ref{trigger_scheme}
(also see Figure~\ref{score_paper_detector_station_concept}).
The analog PMT signals are clipped to a pre-defined amplitude level $u_\mathrm{clip}$ and then summed.
The clipped sum then passes a discriminator set to a threshold level $u_\mathrm{thr}$, with $0<|u_\mathrm{thr}|/|u_\mathrm{clip}|<4$. {The exact value of the ratio $u_\mathrm{thr}/u_\mathrm{clip}$ is a free parameter. In the simulation presented here, this ratio was set to {3.8}}.
A local station trigger is issued when the time-over-threshold of the clipped sum is larger 
than $\Delta \tau$. The clipped sum trigger {prevents false triggers from large signal fluctuations in individual PMTs, therewith suppressing afterpulses,}
NSB photons, and triggers from uncorrelated cosmic-ray muons (see also
Section~\ref{section_trigger_rates}), {allowing a lower station trigger threshold}.
The value of $\Delta \tau$  depends upon the choice of the discriminator. Here, $\Delta \tau$\,=\,7\,ns was used.
At each local station trigger, the data are read out and sent to a central PC farm.
Additionally, all neighbouring stations
which have not issued a trigger are read out as well.
In the simulated array setup (150\,m grid constant), a typical gamma-ray event with 200\,TeV primary energy
triggers of the order of 10 stations. A distribution of photon counts in the simulated array is shown
in Figure~\ref{event_display} for a simulated gamma-ray event at 187\,TeV.
Stations that issue a local trigger in the simulation are additionally marked with a grey circle.

% time synch
\section{HiSCORE simulation results}
\label{hiscore_simulation}
\subsection{Simulation of air showers and detector response}

\paragraph{Simulation software}
% simulation code
Air showers were simulated with CORSIKAv675
%\cite{1998cmcc.book.....H}
\cite{2012ascl.soft02006H}
using the hadronic interaction models QGSJET01c.f \cite{1997NuPhS..52...17K}, and GHEISHA \cite{fesefeldt:1985a},
and the electromagnetic interaction model EGS4.
{The systematic error on the background estimation is due to the limited accuracy of hadronic interaction models,
and to the lack of precise knowledge of the chemical composition of hadrons. While the former will be improved 
with models modified on the basis of LHC data \cite{Pierog:2013qdx}, the latter is one result to be deduced from HiSCORE data
in the future.}
Gamma rays, protons, helium- nitrogen- and iron-nuclei were simulated
in the energy range from 10\,TeV to 5\,PeV following
a power law distribution $dN/dE\propto E^{-1}$ resulting in equal numbers of events per decade.
{Additionally, protons from 5\,TeV to 10\,TeV were simulated for the estimation of the detector trigger rate
(see section~\ref{section_trigger_rates}).}
{Due to our focus to high energies, we require a large Cherenkov light pool.
Thus, the detector array was simulated at an altitude of 0\,m above sea-level.
Simulations at higher altitudes show that a benefit at low energies is only achieved when
using a smaller inter-station spacing, and thus a smaller overall instrumented area
\cite{2010tsra.confE.245H}.}
{Within CORSIKA}, the \texttt{IACT} option \cite{2008APh....30..149B} was used, 
storing Cherenkov photons in spheres of 1\,m radius at sea-level,
each sphere representing one detector station.
{Different geometrical array layouts were simulated.} The {standard} array layout {consists of 484 stations distributed over a regular square grid} as shown in Figure~\ref{event_display}.
% as a simple grid of 22\,$\times$\,22 stations
%with an inter-station spacing of 150\,m,
The standard array covers a total instrumented area of roughly 10\,km$^2$ (3.15\,km side length). Each station in the
standard configuration was simulated based on 8'' PMTs as described above.
Air showers were simulated with uniformly distributed impact position in a square with side length 3.8\,km, corresponding
to a simulated area of 14.44\,km$^2$. The directions were simulated randomly with a uniform density in solid angle 
up to a zenith angle of 30$^\circ$. 
%The data set used for the results presented in this paper contains about
%7\,500 events per particle per decade in energy.

\begin{figure}[ht!]
\centering
\includegraphics[width=\columnwidth]{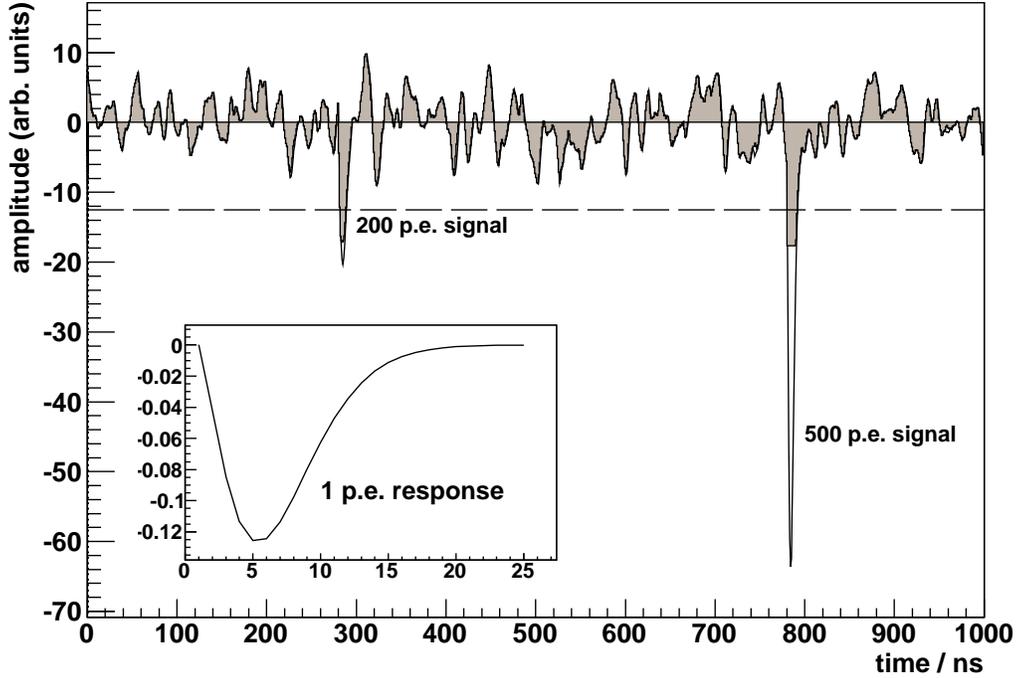}
\caption{The inlay shows a normalized PMT response for a single photoelectron (p.e.) as a function of time (ns).
            The main figure shows a 1$\mu$s long measurement of simulated night sky background (NSB)  (after summing, 
 	 clipping and subtraction of baseline). Added to the NSB are two triggering events from air showers at 
	the level of 200~\,p.e. and 500~\,p.e.:
 Shown are the simple analog sum {(black line)} and the sum of the clipped signals
	 {(filled grey histogram)}
	 used for the trigger decision. {The dashed line indicates the trigger threshold $u_{thr}$ used here.}}
\label{pmt_signals}
\end{figure}

A full detector simulation (\emph{sim\_score}) was implemented
on the basis of the \texttt{IACT} package \cite{2008APh....30..149B}.
At the position of each CORSIKA sphere, a detector station with 4 PMT-channels
is simulated in \emph{sim\_score}.
In detail, \emph{sim\_score} comprises the following elements:
\begin{itemize}
\addtolength{\itemsep}{-0.1cm}
\item Winston cone acceptance tables based on ray-tracing simulations \cite{hampf:phd}.
\item Atmospheric scattering of the Cherenkov photons is calculated using MODTRAN \cite{modtran}.
\item PMT quantum efficiency (wave-length dependent) as in ElectronTubes data sheet for KB9352 (29\,\% at maximum).
\item Overall PMT photoelectron collection efficiency of 90~\%.
\item PMT signal pulse shape (see inlay of Fig.~\ref{pmt_signals}) using a parametrization by
\cite[][see~\ref{pulse_shape_appendix}]{henke:dip}.
\item Station trigger: the clipped sum station trigger as in Figure~\ref{trigger_scheme}.
\item The NSB photon baseline is simulated separately and
      added to the signals at read-out level (pulse shaping and afterpulses are included in this simulation).
\end{itemize}
%
% simulation of NSB event array
%
An interval of 2\,s of NSB noise is simulated as an array of photons uniformly distributed over time, and corresponding to
a constant flux of $3\,\times\,10^{12}$\,photons/(m$^2$ sr s).
%
% NSB response
%
An average quantum efficiency for NSB photons of 0.1 (calculated
using the wavelength dependent quantum efficiency of the PMTs and a spectrum for NSB photons),
and a constant photo electron conversion efficiency of 0.9 were used.
Each photo electron is stored in a read-out array (each array element corresponding to a 1\,ns DRS\,4 cell),
according to the shape and amplitude response described above.
A 1\,$\mu$s interval of simulated NSB noise after subtraction of the average baseline is shown in Figure~\ref{pmt_signals},
along with two signals corresponding to 200 and 500\,photoelectrons (p.e).
The results presented in this paper were obtained using a clipped sum threshold corresponding to 180\,p.e ($u_{thr}$, see Figure~\ref{trigger_scheme}).
%The discriminator threshold is defined in units of peak-p.e..
%One peak-p.e. corresponds to an amplitude of the summed signal of $N \cdot H_{peak}$.
%The conversion of the peak-p.e. value to an actual p.e. value depends on the discriminator response time.
%For a 0\,ns response time, these units are identical. For a 7\,ns response time, 100\,peak-p.e. correspond to 170\,p.e.

\subsection{Trigger rates}
\label{section_trigger_rates}
\begin{figure}[ht!]
\centering
\includegraphics[width=\columnwidth]{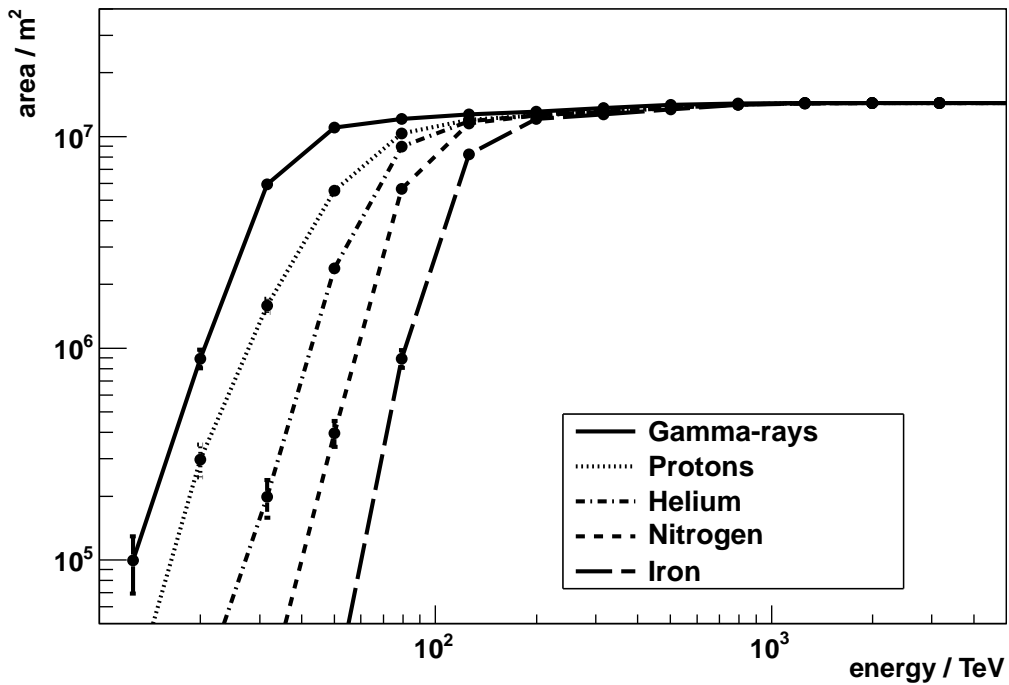}
\caption{Effective areas $A_{eff}$ of the HiSCORE detector at trigger level (1 station trigger)
	 for the five primary particle types simulated in this study. {The standard configuration
	 presented here results in an average number of 3 triggered stations for a 50\,TeV gamma-ray.}}
\label{aeff_trigger}
\end{figure}

{The relevant contributions to the trigger rate are night-sky background (NSB) photons, and cosmic rays. In the following paragraphs,
we present estimations of these rates, and show that the flux of uncorrelated atmospheric muons does not contribute.}
\paragraph{Hadron trigger rates}
The differential trigger rate $R(E)$ of an array can be calculated in the following way:
\begin{eqnarray}
 R(E) &=& \sum\limits_\mathrm{Z=1}^{92} \Phi_Z(E) \cdot \int d\Omega A_\mathrm{eff}(i \in \{\mathrm{p,L,M,H}\},E,\vartheta,\varphi),
\end{eqnarray}
 where the sum runs over the individual cosmic-ray fluxes $\Phi_i(E)$ of 
the elements {as provided by \citet{2003APh....19..193H}} and the effective areas for
cosmic ray nuclei  
have been derived for representative elements:   Helium for the light (L: Z=2-5), 
Nitrogen for the medium (M: Z=6-24), 
and iron for the heavy (H: Z$>$24) groups. 
The effective area  is given as the 
ratio of triggered to simulated events {multiplied} by
the simulated area.  
 The effective area $A_\mathrm{eff}$ depends in principle on the zenith angle $\vartheta$
as well as on the azimuth angle $\varphi$. 
The simulations show however, that up to $\vartheta\approx 25^\circ$, the 
effective area does not vary strongly which justifies a simplified treatment: 
the effective area is assumed to be constant
over $\vartheta$ and the effective solid angle within the constant region
($\vartheta\,<\,25^\circ$) is $\Delta \Omega=0.6$~sr.
%\sout{divided}
In Figure~\ref{aeff_trigger}, $A_\mathrm{eff}(i,E)$ is given for gamma-rays and the 4 hadronic
particle types (H, He, N, Fe - corresponding to protons, light, medium and heavy groups),
requiring only one station to trigger at a single station threshold of 180\,p.e. and using the trigger scheme described in
Figure~\ref{trigger_scheme}.
% effective area trigger level

\begin{figure}[ht!]
  \includegraphics[width=\columnwidth]{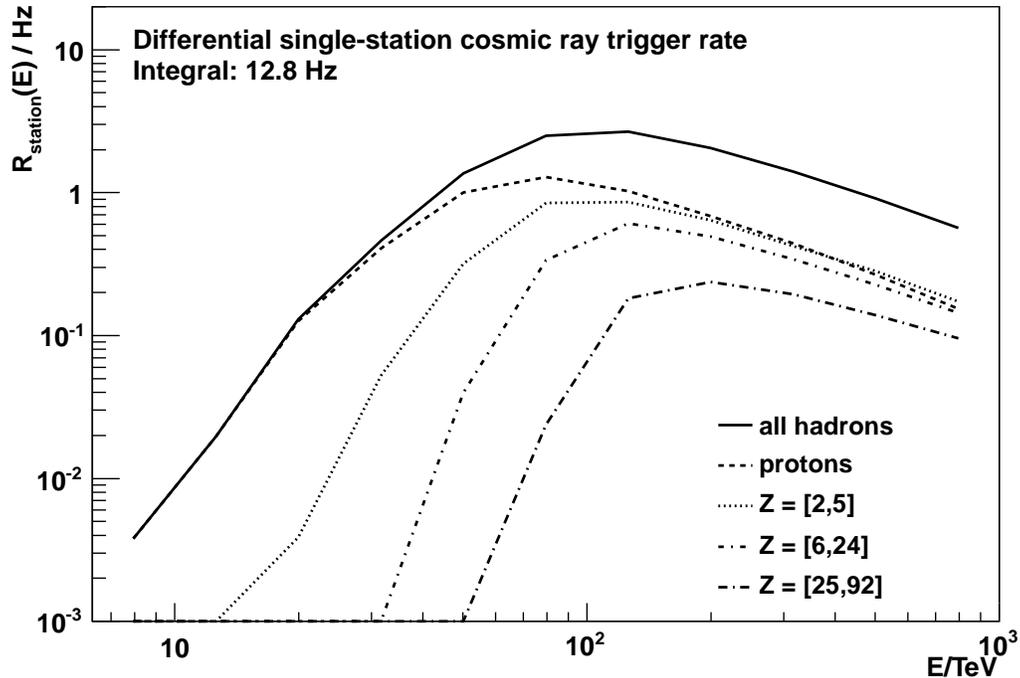}
  \caption{\label{hadron_rate}Differential all-hadron trigger rate for one single HiSCORE station.
	   A local station trigger threshold of 180\,p.e. was used.}
\end{figure}

\begin{figure}[ht!]
  \includegraphics[width=\columnwidth]{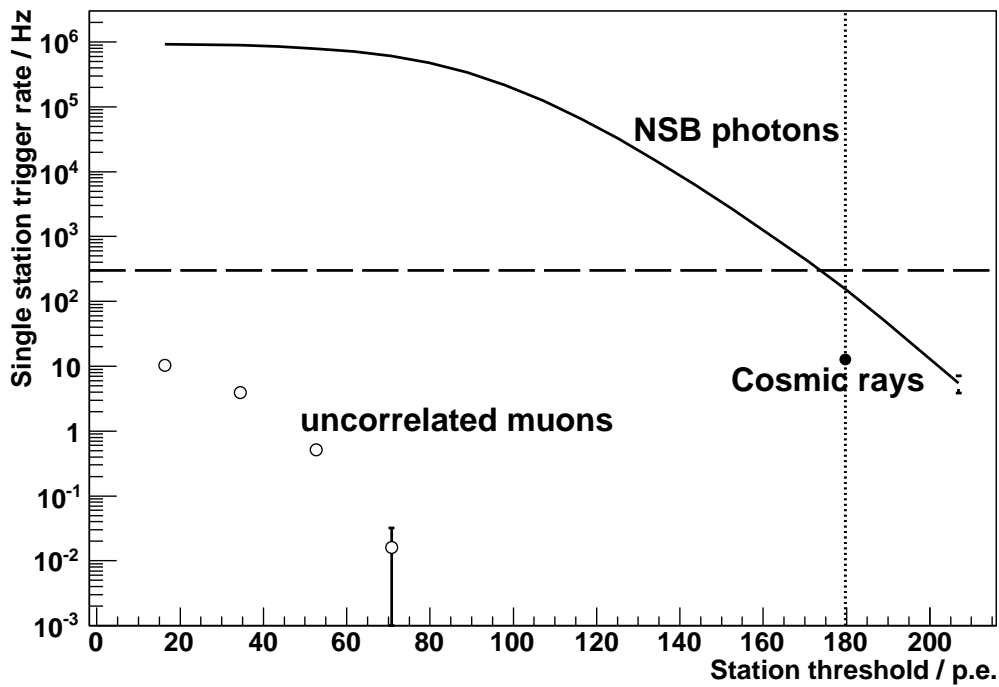}
  \caption{\label{all_rates}Trigger rates as a function of station threshold for muons and NSB photons. For comparison,
   the all-hadron trigger rate is shown as a full circle. For data flow minimization, we aim at a maximum
   trigger rate of 300\,Hz, as indicated by the dashed line.}
\end{figure}

A parametrization of the cosmic ray spectrum \cite{2003APh....19..193H}
is weighted with the corresponding effective areas
to calculate the expected trigger rates for hadronic cosmic ray events.
The resulting trigger rate as a function of energy for an
individual HiSCORE station is shown in Figure~\ref{hadron_rate}.
The integral single-station local trigger rate is found to be 12.8\,Hz.
%When using the first trigger-level only (single local station trigger, 180\,p.e. clipped sum threshold),
The hadron trigger rate of the full 10\,km$^2$-array of 484 stations is  1.77\,kHz
(p: 875\,Hz, L: 505\,Hz, M: 290\,Hz, H: 100\,Hz).
\paragraph{Night-sky background trigger rates}
The night-sky background (NSB) is induced by light sources outside of the atmosphere including direct and scattered starlight, scattered moon light, 
zodiacal light and light produced within the atmosphere including 
air glow and  human-made light sources.
The expected trigger rate induced by NSB photons was simulated using a 
dedicated simulation of a full 4-channel station including
the pulse shaping and trigger scheme. 
Measurements of the NSB level at Fowlers gap in Australia \cite{2011AdSpR..48.1017H}
were used as input for these simulations.
An NSB-induced single-station local trigger rate of 100\,Hz was found 
at a clipped sum threshold of 180\,p.e.,
demonstrating that the single-station trigger rate is clearly dominated by NSB photons (see Figure~\ref{all_rates}).
A reduction of the data flow could be achieved when also using a second trigger level,
requiring coincident ($\mu$s-window) triggers of 2 neighbouring stations.
This could be realized on software level and is not necessary on hardware level as long
as the total data flow from single-station triggers can be handled {(e.g. using a central PC farm)}.
Alternatively, slightly increasing the station discriminator threshold to 190\,p.e.~results in an NSB trigger rate of 50\,Hz
without significantly affecting the sensitivity{, and only raising the energy threshold (linearly) by 5\,\%}.

\paragraph{Uncorrelated cosmic ray muon trigger rate}
The overall flux of vertical incidence atmospheric muons measured at sea
level is 160\,particles\,m$^{-2}$s$^{-1}$ \cite{2010PDG}.
Taking into account the detector properties, the
total average signal from one muon in one station is of the order of 100\,p.e.
%i.e. significantly below the threshold for the clipped sum of 180\,p.e. used here.
The trigger rate for uncorrelated muons as shown in Fig.~\ref{all_rates} was 
calculated with the full simulation chain (CORSIKA \& \emph{sim\_score}),
using different clipped sum thresholds. 
As can be seen from Figure~\ref{all_rates},
the muon trigger rate is found to be negligible in comparison 
with the rate of NSB photon induced triggers. The rate was calculated for a 
trigger setup with a clipped sum threshold of 180\,p.e. and 
$\Delta \tau = 7$\,ns.

%at reconstruction level ($\ge$\,3\,Stations and reconstruction cuts).
{
\subsection{Air shower reconstruction}
\label{section:air_shower_reconstruction}
Air shower reconstruction algorithms for HiSCORE were introduced in \cite{2013NIMPA.712..137H,hampf:phd}.
The arrival direction is reconstructed using an analytical model for the arrival time of Cherenkov photons at the detector
stations. The primary particle energy is reconstructed from the light density value at a fixed distance from the shower core,
as reconstructed from the lateral density function of a shower event.
As shown in \cite{2013NIMPA.712..137H}, the standard array configuration achieves an angular resolution of
0.25$^\circ$ at 100\,TeV and 0.1$^\circ$ at 1\,PeV. The achieved energy resolution is 20\,\%
at 100\,TeV, improving to better than 10\,\% at 1\,PeV.
The resolution of the height of the shower maximum is 70\,g/cm$^2$ at 50\,TeV and reaches 40\,g/cm$^2$ at 1\,PeV.
These numbers, even though obtained with very basic reconstruction algorithms,
are comparable to the energy and shower maximum resolutions of the Tunka-133 array (15\,\% energy-, and 25\,g/cm$^2$ shower maximum resolution at higher energies) \cite{2012NIMPA.692...98B}.
Primary particle identification is done on the basis of the reconstructed energy
and shower height.
At the energy threshold and up to 100\,TeV, the gamma-hadron separation
is inefficient (quality factor of 1.0). Beyond 100\,TeV, the quality factor gradually improves to 2.0 at 1\,PeV.
A comprehensive discussion can be found in \cite{hampf:phd}.
The quality of spectroscopic measurements of the all-particle cosmic ray spectrum will depend on the energy resolution
and the total effective area of the instrument. At 10\,\% energy resolution and an instrumented area of 100\,km$^2$,
HiSCORE will allow a high accuracy for spectroscopic reconstruction of cosmic rays around the knee region and beyond.
}
\subsection{Sensitivity to gamma-rays}
\label{section:gamma_sensitivity}
% Sensitivity at reco level
The point-source survey sensitivity (for a definition see below) to gamma-rays
was calculated on the basis of the simulation {and reconstruction} described above.

\begin{figure}[ht!]
  \includegraphics[width=\columnwidth]{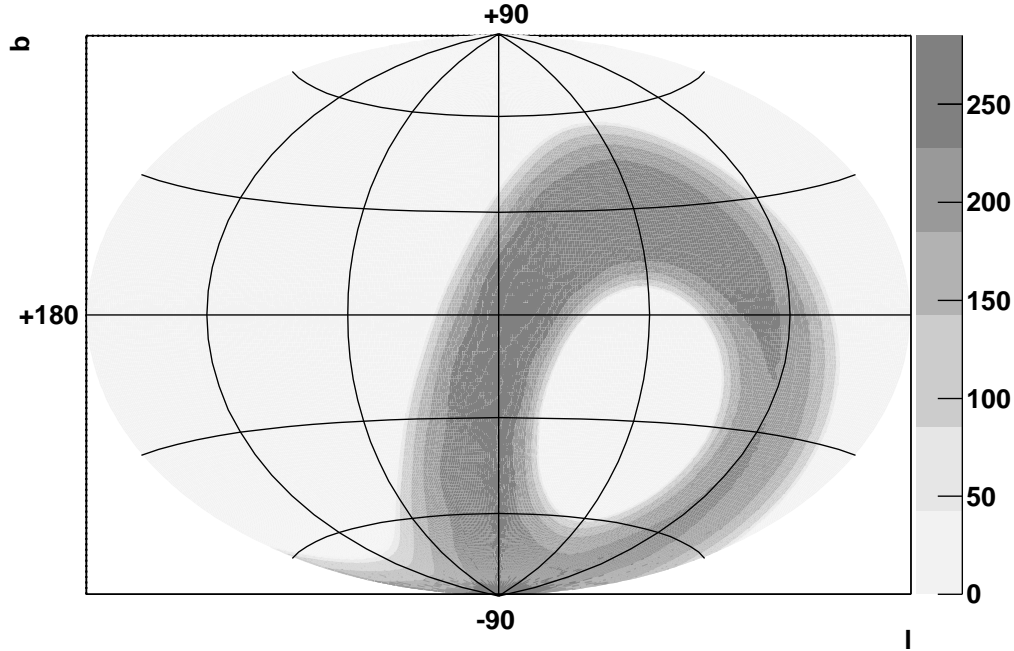}
  \caption{\label{exposure}Exposure in hours for one year of data taking at a southern site
(31$^\circ$ southern geographical latitude),
using only events within 25$^\circ$ half opening angle. An area of one $\pi$\,sr is covered for more than 200\,h
within one observation year.}
\end{figure}

\begin{figure}[ht!]
  \includegraphics[width=\columnwidth]{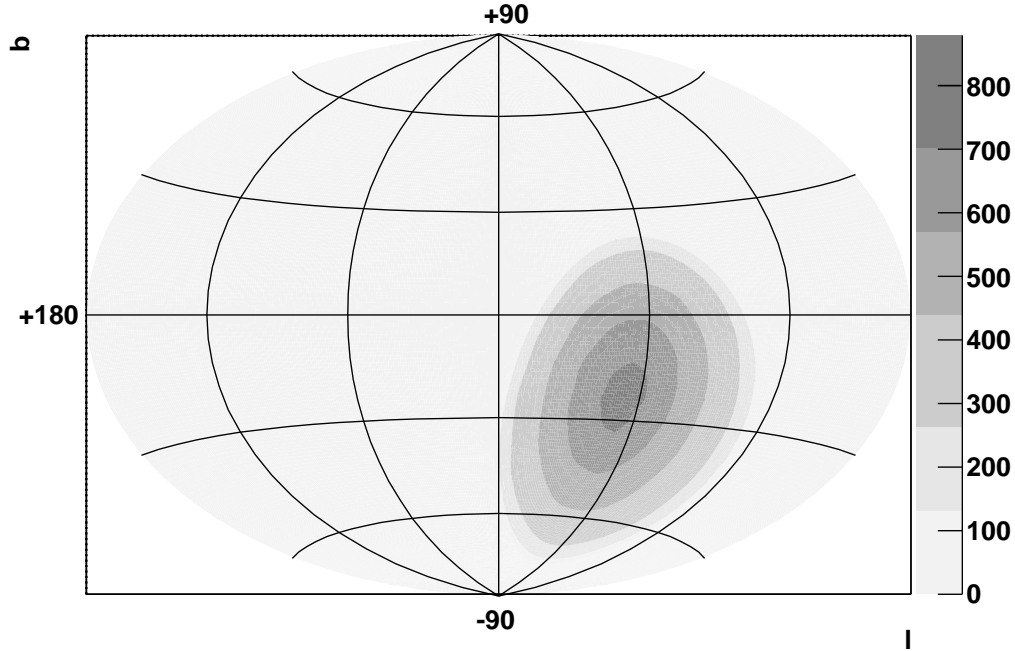}
  \caption{\label{exposure_tilted}Exposure in hours for one year of data taking in a 30$^\circ$ \emph{tilted south mode},
at a southern site (31$^\circ$ southern geographical latitude), using only events within 25$^\circ$ half opening angle.}
\end{figure}

\begin{figure*}[ht!]
\centering
\includegraphics[width=\textwidth]{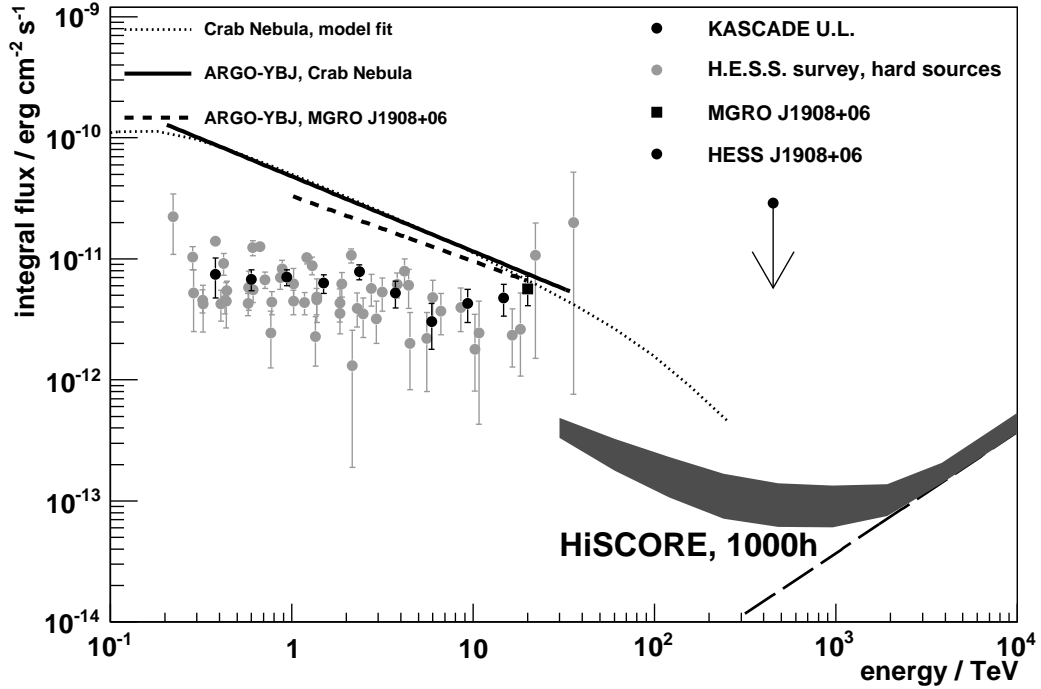}
\caption{
                HiSCORE point-source \emph{survey} sensitivity {to gamma-rays}.
		{The chosen 1000\,h of observation time correspond to 5\,years of standard mode operation
		and roughly 1.4\,years of tilted mode operation.}
		The dotted line represents a model fit to data on the Crab Nebula from different experiments \cite{2010A&A...523A...2M}.
		The best fit results from ARGO-YBJ on the Crab Nebula are show as a solid line \cite{2013ApJ...779...27B}.
                Also shown for comparison are VHE data from the first H.E.S.S. survey of the inner Galactic plane
		\cite{2006ApJ...636..777A}, the Milagro source MGRO\,J1908+06
		(Milagro data \cite{2007ApJ...664L..91A}, H.E.S.S. data \cite{2009A&A...499..723A},
		ARGO-YBJ data \cite{2012ApJ...760..110B}), and an upper limit from KASCADE \cite{2004ApJ...608..865A}.
	 }
\label{sensitivity_sources}
\end{figure*}

%\paragraph{Observation time and modes}
%
HiSCORE always operates in \emph{survey mode}, i.e. all objects which are visible during darkness time within the 
visible cone of the instrument are observed. 
In the standard observation mode, the optical axis of the detector stations points to the zenith.
The observation time for any given position in the sky was calculated for an observation site at 31$^\circ$ southern geographical latitude.
The accumulated exposure time for a celestial position is derived using an acceptance cut taking into account 
only events within 25$^\circ$ half opening angle \cite{2010tsra.confE.245H}.
While the region of the sky covered is restricted by the detector site,
the large effective field of view of 0.6\,sr leads to a total sky coverage of $\pi$\,sr
in one year at an exposure depth ranging from 200\,h to 283\,h, as illustrated in
Figure~\ref{exposure}.
For comparison, the first survey of the inner Galaxy with H.E.S.S. covered a
total solid angle of the order of 0.1\,sr at an exposure level of the order of 10\,h.
{The amount of survey time that can be allocated with a pointing instrument (IACT) depends on the
time dedicated for deep exposure of individual objects, which might have precedence over a survey. The advantage of
IACTs is their flexibility when determining the exact region to survey.
The situation is different for HiSCORE. The region that is surveyed is defined by the region covered by the
(fixed) instrument within a year. We plan to improve this situation using a reorientation of the detectors along
their north-south axis} (tilting), allowing to access different regions of the sky.
%Different regions of the sky can be accessed by a reorientation of the optical detector axis (tilting).
This way, after a given operation time in standard observation mode (e.g. few years without tilting),
the total sky coverage can be increased,
by consecutively tilting the detector stations to the south (\emph{tilted south observation mode})
and to the north for additional operation periods of a few years.
{For illustration, using a tilting angle of 30$^\circ$ towards the south (for a southern observation site),
the sky covered is extended to the region shown in Figure~\ref{exposure_tilted},
with a significantly deeper exposure per source within this region.}
%\delete{For a (conservative) calculation of the sensitivity, only the standard configuration (no tilting) was considered.}
%Furthermore, no envisagable extensions or hybrid detector parts were taken into account here (e.g. muon detectors, imaging
%telescopes).
%
%
%
{We used a total observation time of 1000\,h, corresponding to 5\,years of standard mode operation and roughly 1.4\,years of tilted mode operation.}
A minimum of 5\,$\sigma$ detection significance\footnote{The significance is calculated using eqn. 9 of \cite{1983ApJ...272..317L}} and 50 gamma-rays were required to define the flux sensitivity.
The background numbers were calculated in an analogous way as described in
Sect.~\ref{section_trigger_rates} using however effective areas 
after {quality cuts} \cite[][]{2013NIMPA.712..137H}. 
The resulting point-source sensitivity {for 100\,km$^2$ instrumented area is shown in Figure~\ref{sensitivity_sources}.
The dark shaded area shows the range of sensitivities achieved when using pessimistic
(upper bound) and optimistic (lower bound) assumptions \cite[also see][]{2013NIMPA.712..137H}.
The pessimistic bound was obtained by using a conservative alpha factor
(ratio of solid angles of source to background measurements in Eq. 9 of \citet{1983ApJ...272..317L})
of $\alpha=1$ and an angular gamma-ray point source cut
(the size of the source region around a test position) efficiency of 0.68.
In the optimistic scenario, $\alpha<<1$.
Additionally, the gamma-ray efficiency of the point source cut was set to 1
in the background free regime, i.e. above 2.1\,PeV.
Also shown in Figure~\ref{sensitivity_sources} are} representative flux measurements
from H.E.S.S. \cite{2006ApJ...636..777A},
ARGO-YBJ \cite{2013ApJ...779...27B},
Milagro \cite{2007ApJ...658L..33A}, and an upper limit from KASCADE
\cite{2004ApJ...608..865A}.

At the energy threshold, the sensitivity of HiSCORE is mainly limited by the angular resolution
and secondarily by the gamma-hadron separation {(See Section~\ref{section:air_shower_reconstruction})}.
At the upper energy end, the sensitivity is limited by count statistics
and depends linearly on the inverse of the product of total detector area and the exposure time,
{forming the background free regime (straight black dashed line in Figure~\ref{sensitivity_sources}).
In the central energy regime, the sensitivity is limited by the gamma-hadron separation.
Here, we only show the sensitivity for the standard layout. Improvements of the angular resolution
or to the gamma-hadron separation (large PMTs, graded array, muon detectors, see Section~\ref{alternative_configurations})
are not included.
Such detector enhancements will cause the background free regime to extend to lower energies, effectively pushing down the
sensitivity curve toward the straight black dashed line.
}
With HiSCORE, it will be possible to study the continuation of the spectra 
of known Galactic sources up to several 100\,TeV.
In this context, it will be important to investigate whether some of these sources could be cosmic ray
pevatrons \cite{2007ApJ...665L.131G}, \cite[also see][]{2011AdSpR..48.1935T,2012NIMPA.692..246T}.

\begin{figure*}[ht!]
\centering
\includegraphics[width=\textwidth]{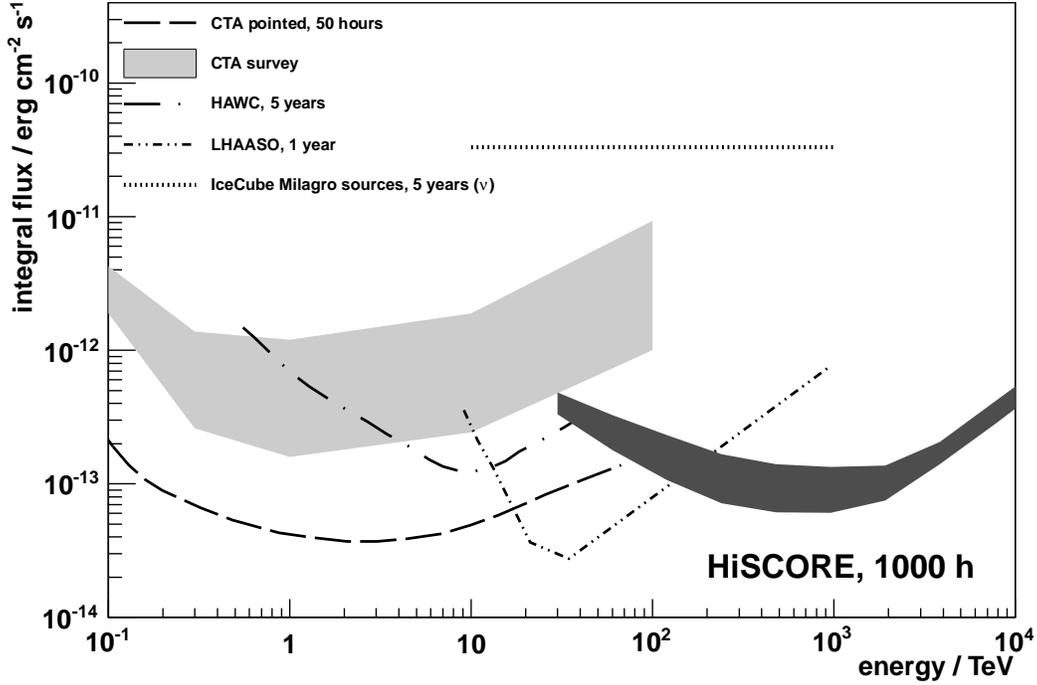}
\caption{
                HiSCORE point-source \emph{survey} sensitivity {after 1000\,h of exposure time}
		in comparison with sensitivities from other experiments.
		{The chosen 1000\,h of observation time correspond to 5\,years of standard mode operation
                and roughly 1.4\,years of tilted mode operation.}
%                Also shown for comparison are VHE data from the first H.E.S.S. survey of the inner Galactic plane
%		\cite{2006ApJ...636..777A}, the Milagro source MGRO\,J1908+06
%		(Milagro data \cite{2007ApJ...664L..91A} and H.E.S.S. data \cite{2009A&A...499..723A}),
%		and an upper limit from KASCADE \cite{2004ApJ...608..865A}.
%		%\cite{2007ApJ...658L..33A}.
		For direct comparison,
		the point-source \emph{survey} sensitivities of CTA \cite{2013APh....43..317D},
		of a search for neutrinos by 
		IceCube \cite[Milagro source stacking,][]{2009APh....31..437G}, 
		HAWC \cite[][]{2005AIPC..745..234S},
		and LHAASO \cite[][]{2010ChPhC..34..249C,Cui201486} (adapted to a minimum of 50 gamma-rays)
		are shown (all three northern hemisphere).
                For reference, also the 50-hour \emph{pointed-observation} sensitivity of CTA \cite{2011ExA....32..193A} is given.
	 }
\label{sensitivity_experiments}
\end{figure*}

A comparison of the senstivity of the HiSCORE standard configuration to point-source sensitivities
of other gamma-ray experiments (CTA \cite{2011ExA....32..193A,2013APh....43..317D}, HAWC \cite{2005AIPC..745..234S},
and LHAASO \cite{2010ChPhC..34..249C,Cui201486}, the latter {being valid for an exposure of one calendar year, and}
adapted to a minimum statistics requirement of 50 events) is shown in Figure~\ref{sensitivity_experiments}.
{The sensitivity of IceCube for
detecting a $5~\sigma$ excess of neutrinos after 4.5~years of observation 
from stacked gamma-ray sources as explained in \citet{2009APh....31..437G} is also included.
Here, the underlying assumption is that the entire gamma-ray emission is of hadronic origin.}
For comparison of sensitivities, it is important to note that the sensitivity shown for HiSCORE is the
point-source \emph{survey sensitivity for $\pi$\,sr of the sky}. While the shown sensitivities for HAWC, LHAASO, and IceCube are
also valid for surveys, the CTA sensitivity is given for pointed observations of 50\,h in a small
field of view of the order of $\pi / 100$\,sr (dashed line), for a survey of $\pi$\,sr (upper bound of grey area) and a survey
of the Galactic plane (lower bound of grey area).
{Furthermore, it has to be noted that the total observation time available per year to HiSCORE
depends on the observation mode (see above).}

\section{Alternative array configurations}
\label{alternative_configurations}
{
For comparison to the standard array geometry and station configuration presented above, two further configurations were
tested using the same simulation framework.
\paragraph{Alternative configuration (a) -- 12'' PMTs}
This configuration consists of
a standard geometry array configuration equipped with 12'' PMTs and correspondingly larger Winston cones,
resulting in a 2.25-fold increase in light collection area per station.
%Additionally, the 12'' PMTs were assumed to have a 10\% higher quantum efficiency as compared to the ones used in the standard
%configuration.
As can be expected, the usage of larger PMTs allows to detect lower Cherenkov photon densities on the ground, therewith
lowering the energy threshold of the detector array for air showers, i.e. maximizing the effective area at low energies.
For this array configuration,
a higher individual station threshold (factor 1.5) was used to limit the expected data rate to a level comparable to
the expected rate for the standard array.
\paragraph{Alternative configuration (b) -- graded array}
A circular graded array layout of 493 stations
with a dense core and inter-station distances gradually decreasing towards the edge
of the array is shown in Figure~\ref{graded_array}.
At a comparable cost in number of stations, this layout results in improved effective areas at low energies
(small inter-station distances in the array core) and at the same time optimizes the area at large energies due to the
much larger area covered by the total array.
%(see Figure~\ref{effective_area_comparison}).
\begin{figure}[!ht]
 \centering
 \includegraphics[width=\columnwidth]{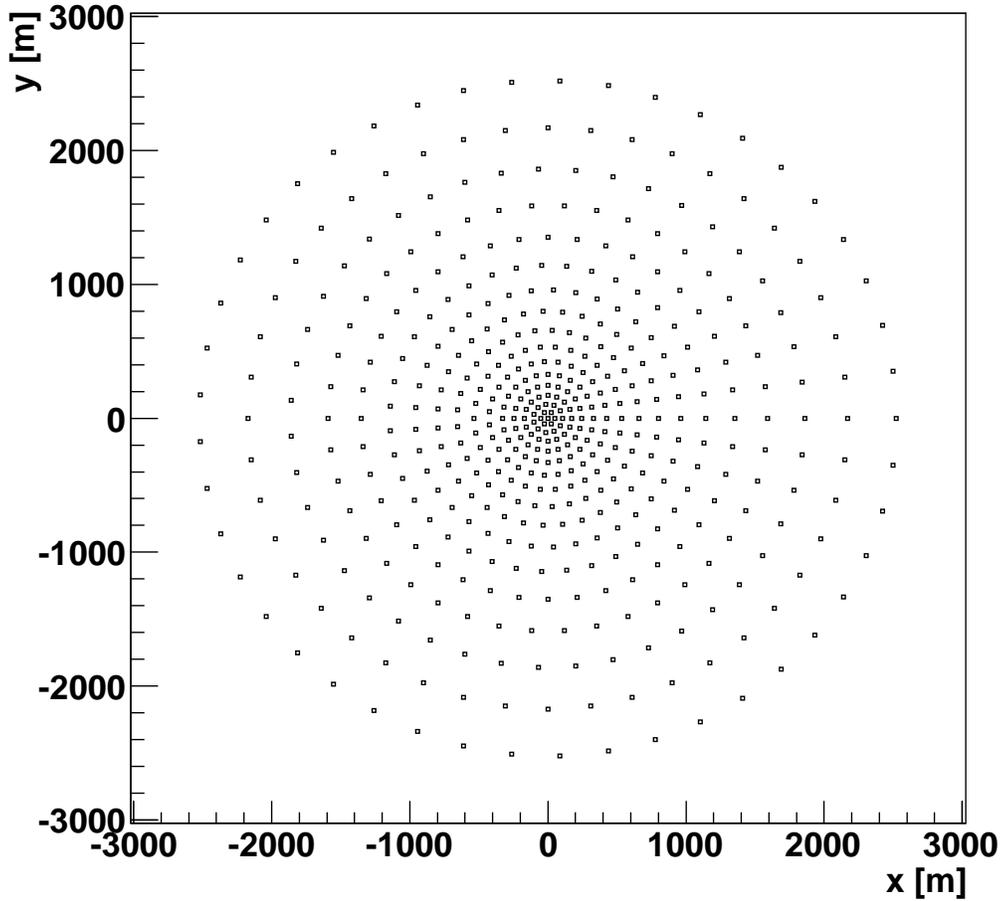}
 \caption{Graded array layout (b). Using a comparable number of stations (493 as compared to 484 for the standard
          layout), the graded geometry optimizes the detection and reconstruction efficiency at low energies in the
          dense core of the array, and optimizes the sensitivity to large energies due to the larger area covered.
 \label{graded_array}}
\end{figure}
\paragraph{Performance of alternative arrays}

% effective areas
The resulting effective gamma-ray trigger areas for these configurations are shown in Figure~\ref{effective_area_comparison}.
\begin{figure}[!ht]
 \centering
 \includegraphics[width=\columnwidth]{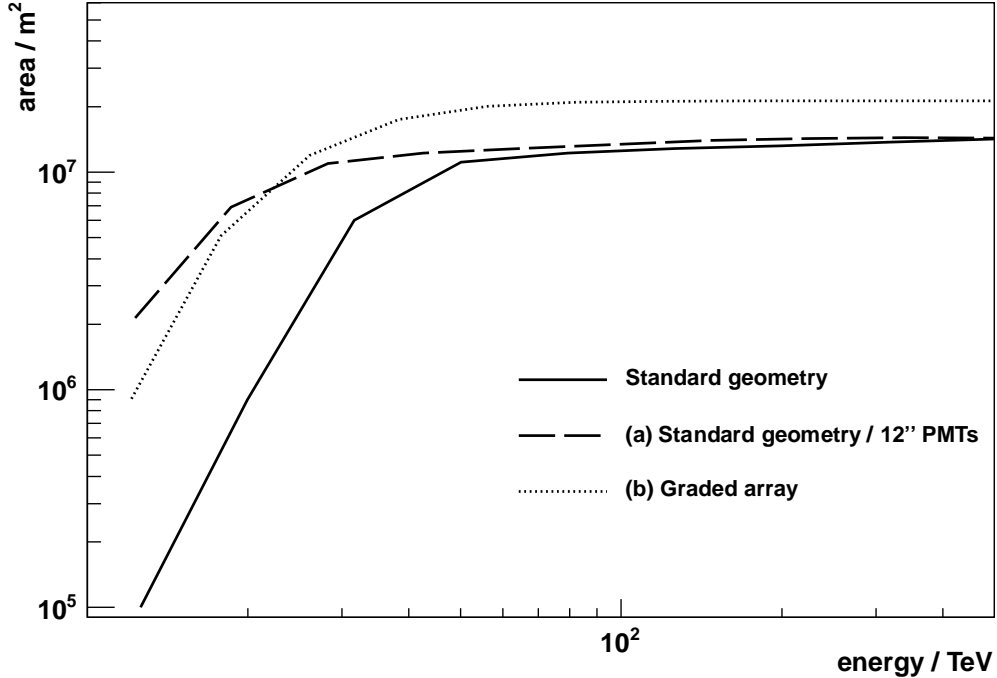}
 \caption{Effective gamma-ray areas at trigger level for the alternative detector configurations (a) standard geometry using large 12'' PMTs, and (b) graded array layout. The effective area for the standard configuration is shown
for comparison (solid line).
 \label{effective_area_comparison}}
\end{figure}
The described maximization of the effective area is clearly visible
for configuration (a) at low energies and configuration (b) over the full energy range.

A comparison of the angular resolutions for both alternative configurations and the standard configuration is
shown in Figure~\ref{angular_resolutions}.
% angres (a) low E
Configuration (a) shows an improvement of the angular resolution primarily at low energies, where the larger
light collecting area of the 12'' PMT stations
improves the signal to noise ratio in each station
and, more importantly, increases the number of stations per shower event,
the stations being more sensitive to low light levels, i.e. larger core distances (also see Figure~\ref{score_paper_lateral}).
% angres (a) high E
Towards high energies the latter advantage becomes less significant because the relative increase
in number of stations is less strong for large shower events.
% angres (b) low E
The dense core of the graded array layout (b) also leads to a larger number of stations at lower energies
(a larger number of stations being located inside the inner Cherenkov light pool, also see Figure~\ref{score_paper_lateral}),
i.e. an improved angular resolution as compared to the standard configuration.
% angres (b) high E
However, at higher energies (E\,$>$\,330\,TeV) a slight deterioration as compared to the standard configuration is seen.
This can be explained by the fact that most events at higher energies are detected by the low-density part of the array,
with fewer stations per shower event than in the standard configuration.
\begin{figure}[!ht]
 \centering
 \includegraphics[width=\columnwidth]{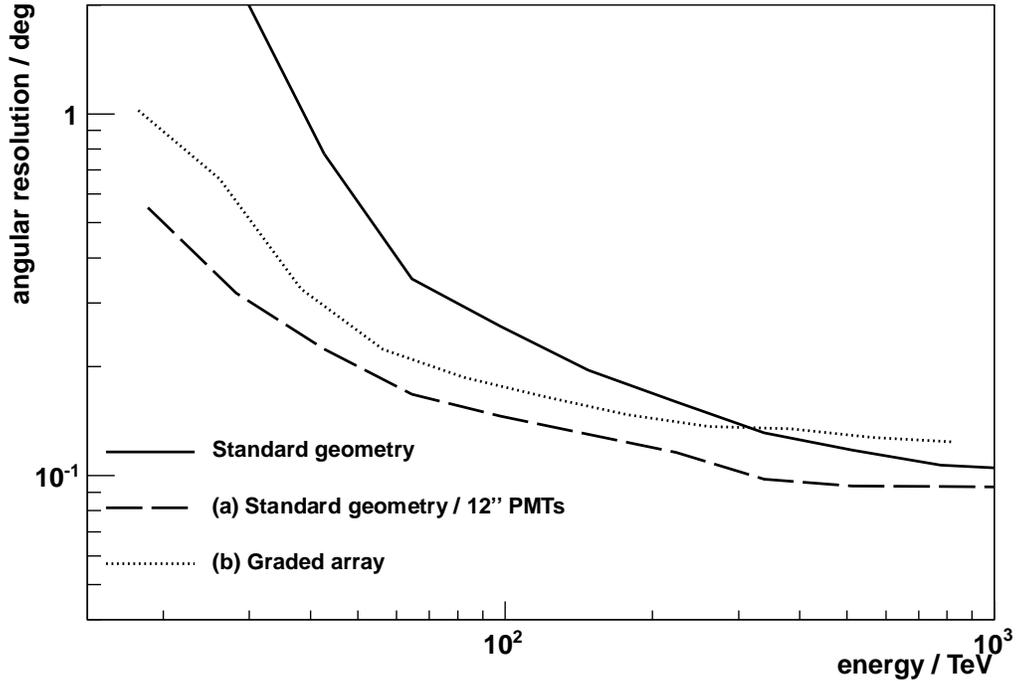}
 \caption{Angular resolution (1\,ns time jitter) {for gamma-rays using} the standard configuration, and the alternative configurations (a) and (b).
 \label{angular_resolutions}}
\end{figure}

Adaptations of the reconstruction (especially gamma-hadron separation)
to different detector configurations are work in progress and
will be the subject of a future publication.
Quite obviously, a combination of a graded array also equipped with large PMTs would further improve
the performance as compared to configurations (a) and (b) alone.
Here, we aim at showing the effect of both alterations to the standard layout separately, for clarity.
}
\section{Outlook}
\label{hiscore_outlook}
More complex array layouts, such as combining partly large PMTs with a graded design
or different variants of graded designs, are under study.
Moreover, for a further improvement at low energies, a cell solution is envisaged
in which the four individual PMT channels are separated by 5--10\,m, building small sub-arrays (cells).
Such cells could provide a better sampling of the central part of the LDF, and better reconstruction
(thus sensitivity) at low energies (10\,TeV to 100\,TeV).
Simulations have shown that a cell solution can lower the energy threshold (defined
as the energy at which the effective area reaches 50\,\% of the instrumented area)
considerably \cite{2010tsra.confE.245H}.
%
%% Szintillator material
An extension of the concept we studied so far is to equip the underside of each
sliding lid of the station boxes with scintillator material.
With such a setup, continuation of data taking during daytime would allow to use
HiSCORE as a very large charged particle shower front sampling array,
effectively increasing the duty cycle by a factor of 9.
Finally, we are also studying the benefits of a combination of the non-imaging technique simulated here with the
imaging technique.

\section{Summary}
\label{hiscore_summary}
We have described the HiSCORE concept for a large-area wide-angle air shower experiment,
{based on an array of non imaging light collecting detector stations}.
A comprehensive simulation of the detector was performed including 
all relevant components (atmosphere, light-collection with Winston cones,
photomultiplier, pulse shaping with afterpulses, station trigger).
The resulting effective areas for various primary particles and the expected trigger rates for background
%(charged cosmic rays, muons, night sky background)
have been calculated for different assumptions on the trigger threshold. 
The complete Monte Carlo simulation of the HiSCORE detector concept shows that such a non-imaging air-Cherenkov detector 
will be sufficiently sensitive to survey a large fraction ($\pi~\mathrm{sr}$) of the sky for gamma-ray sources above
{10--50\,TeV (depending on the final array layout)}
at an energy flux level of a few $10^{-13}$~ergs/cm$^2$s. This sensitivity is comparable to the planned next-generation
Cherenkov telescope array (CTA) at lower energies, effectively extending the sensitive energy range into the UHE gamma-ray regime.
Furthermore, HiSCORE will provide high-statistics measurements of cosmic ray spectra and composition
above 100\,TeV primary energy,
covering the energy range of transition between Galactic and extragalactic origin of cosmic rays,
and up to 10$^{18}$\,eV.

An engineering array with $1-2~\mathrm{km}^2$ is planned for deployment {2014/2015}, aiming at proof-of-principle measurements and first
physics results.
%\cite[see e.g.][]{}.

\section*{Acknowledgements}
We gratefully acknowledge valuable discussions on physics and hardware aspects
with F. Aharonian, R. Antonini, N. Budnev, E. Lorenz, B.K. Lubsandorzhiev, R. Mirzoyan,
G.P. Rowell, F. Schr\"oder, T. Schweitzer, and M. Teshima.  
This research has made use of NASA's Astrophysics Data System Bibliographic Services.
We acknowledge the support of the Helmholtz Russia Joint Research Group (HRJRG-303).
We acknowledge funding for the initiation of international cooperation (TL 51-3) by the \emph{Deutsche
Forschungsgemeinschaft}, DFG, and the Russian RFFI (13-02-12095).
Furthermore, we acknowledge funding provided by the \emph{Helmholtz Ge\-mein\-schaft} in the framework of the
Helmholtz Alliance for Astroparticle Physics (HAP).
\appendix
\section{Pulse shape parametrization}
\label{pulse_shape_appendix}
The single photo-electron response of a PMT is determined by the pulse shape as a function of time,
and a probability distribution of the amplitude response including afterpulses.
Here, we used a probability for a response amplitude at 4\,p.e. of 10$^{-4}$.
The normalized pulse shape ($R_{shape}(t)$)
used in this work {was taken from \citet{henke:dip}. This shape was used to describe the response of AIROBICC, which was
using the same PMT as envisaged for HiSCORE. The pulse shape }
is shown as inlay in Figure~\ref{pmt_signals}. It can be described as
\begin{equation}
 \label{pulseshape}
 R_{shape}(t) = D \cdot t'^a \cdot e ^{\left(-b t'\right)^c}
\end{equation}
where $D$ is a normalisation constant, $a = 1.25$, $b = 0.0414$, $c = 1.48$,
and the time t {(in ns)} is included in $t' = \frac{7}{4} t/1\,ns$.
{The response function $R_{shape}(t)$ is a parametrization of the normalized,
dimension-less single photo-electron response function.
The parameters a, b, and c are dimension-less as well.}

%% References


\section*{References}
\bibliography{tluczykont_hiscore_r3}

\begin{thebibliography}{48}
\expandafter\ifx\csname natexlab\endcsname\relax\def\natexlab#1{#1}\fi
\providecommand{\bibinfo}[2]{#2}
\ifx\xfnm\relax \def\xfnm[#1]{\unskip,\space#1}\fi
%Type = Article
\bibitem[{{Antoni} et~al.(2004){Antoni}, {Apel}, {Badea}
  et~al.}]{2004ApJ...608..865A}
\bibinfo{author}{T.~{Antoni}}, \bibinfo{author}{W.~D. {Apel}},
  \bibinfo{author}{A.~F. {Badea}}, et~al.,
\newblock \bibinfo{title}{{Search for Cosmic-Ray Point Sources with KASCADE}},
\newblock \bibinfo{journal}{The Astrophysical Journal} \bibinfo{volume}{608}
  (\bibinfo{year}{2004}) \bibinfo{pages}{865}.
%Type = Article
\bibitem[{Kotera and Olinto(2011)}]{2011ARA&A..49..119K}
\bibinfo{author}{K.~Kotera}, \bibinfo{author}{A.~V. Olinto},
\newblock \bibinfo{title}{The astrophysics of ultrahigh energy cosmic rays},
\newblock \bibinfo{journal}{Annual Review of Astronomy and Astrophysics}
  \bibinfo{volume}{49} (\bibinfo{year}{2011}) \bibinfo{pages}{119}.
%Type = Article
\bibitem[{{THEMISTOCLE Collaboration} et~al.(1993){THEMISTOCLE Collaboration},
  {Baillon}, {Behr}, {Danagoulian} et~al.}]{1993APh.....1..341T}
\bibinfo{author}{{THEMISTOCLE Collaboration}}, \bibinfo{author}{P.~{Baillon}},
  \bibinfo{author}{L.~{Behr}}, \bibinfo{author}{S.~{Danagoulian}}, et~al.,
\newblock \bibinfo{title}{{Gamma ray spectrum of the Crab nebula in the multi
  {T}e{V} region k20}},
\newblock \bibinfo{journal}{Astroparticle Physics} \bibinfo{volume}{1}
  (\bibinfo{year}{1993}) \bibinfo{pages}{341}.
%Type = Article
\bibitem[{{Karle} et~al.(1995){Karle}, {Merck}, {Plaga}
  et~al.}]{1995APh.....3..321K}
\bibinfo{author}{A.~{Karle}}, \bibinfo{author}{M.~{Merck}},
  \bibinfo{author}{R.~{Plaga}}, et~al.,
\newblock \bibinfo{title}{Design and performance of the angle integrating
  {C}erenkov array \mbox{AIROBICC}},
\newblock \bibinfo{journal}{Astroparticle Physics} \bibinfo{volume}{3}
  (\bibinfo{year}{1995}) \bibinfo{pages}{321}.
%Type = Article
\bibitem[{{Fowler} et~al.(2001){Fowler}, {Fortson}, {Jui}
  et~al.}]{2001APh....15...49F}
\bibinfo{author}{J.~W. {Fowler}}, \bibinfo{author}{L.~F. {Fortson}},
  \bibinfo{author}{C.~C.~H. {Jui}}, et~al.,
\newblock \bibinfo{title}{{A measurement of the cosmic ray spectrum and
  composition at the knee}},
\newblock \bibinfo{journal}{Astroparticle Physics} \bibinfo{volume}{15}
  (\bibinfo{year}{2001}) \bibinfo{pages}{49}.
%Type = Article
\bibitem[{{Berezhnev} et~al.(2012){Berezhnev}, {Besson}, {Budnev}
  et~al.}]{2012NIMPA.692...98B}
\bibinfo{author}{S.~F. {Berezhnev}}, \bibinfo{author}{D.~{Besson}},
  \bibinfo{author}{N.~M. {Budnev}}, et~al.,
\newblock \bibinfo{title}{{The Tunka-133 EAS Cherenkov light array: Status of
  2011}},
\newblock \bibinfo{journal}{Nuclear Instruments and Methods in Physics Research
  A} \bibinfo{volume}{692} (\bibinfo{year}{2012}) \bibinfo{pages}{98}.
%Type = Article
\bibitem[{{Dyakonov} et~al.(1986){Dyakonov}, {Knurenko}, {Kolosov}
  et~al.}]{1986NIMPA.248..224D}
\bibinfo{author}{M.~N. {Dyakonov}}, \bibinfo{author}{S.~P. {Knurenko}},
  \bibinfo{author}{V.~A. {Kolosov}}, et~al.,
\newblock \bibinfo{title}{{The use of Cherenkov detectors at the Yakutsk cosmic
  ray extensive air shower array}},
\newblock \bibinfo{journal}{Nuclear Instruments and Methods in Physics Research
  A} \bibinfo{volume}{248} (\bibinfo{year}{1986}) \bibinfo{pages}{224}.
%Type = Article
\bibitem[{{Huege} et~al.(2012){Huege}, {Apel}, {Arteaga}, {Asch}
  et~al.}]{2012NIMPA.662S..72H}
\bibinfo{author}{T.~{Huege}}, \bibinfo{author}{W.~D. {Apel}},
  \bibinfo{author}{J.~C. {Arteaga}}, \bibinfo{author}{T.~{Asch}}, et~al.,
\newblock \bibinfo{title}{{The LOPES experiment - Recent results, status and
  perspectives}},
\newblock \bibinfo{journal}{Nuclear Instruments and Methods in Physics Research
  A} \bibinfo{volume}{662} (\bibinfo{year}{2012}) \bibinfo{pages}{72}.
%Type = Article
\bibitem[{{Gorham} et~al.(2008){Gorham}, {Lehtinen}, {Varner}
  et~al.}]{2008PhRvD..78c2007G}
\bibinfo{author}{P.~W. {Gorham}}, \bibinfo{author}{N.~G. {Lehtinen}},
  \bibinfo{author}{G.~S. {Varner}}, et~al.,
\newblock \bibinfo{title}{{Observations of microwave continuum emission from
  air shower plasmas}},
\newblock \bibinfo{journal}{Physical Review D} \bibinfo{volume}{78}
  (\bibinfo{year}{2008}) \bibinfo{pages}{032007}.
%Type = Article
\bibitem[{{{\v S}m{\'{\i}}da} et~al.(2013){{\v S}m{\'{\i}}da}, {Baur},
  {Bertaina} et~al.}]{2013EPJWC..5308010S}
\bibinfo{author}{R.~{{\v S}m{\'{\i}}da}}, \bibinfo{author}{S.~{Baur}},
  \bibinfo{author}{M.~{Bertaina}}, et~al.,
\newblock \bibinfo{title}{{Observation of microwave emission from extensive air
  showers with CROME}},
\newblock \bibinfo{journal}{European Physical Journal Web of Conferences}
  \bibinfo{volume}{53} (\bibinfo{year}{2013}) \bibinfo{pages}{8010}.
%Type = Article
\bibitem[{{CTA~Consortium} et~al.(2011){CTA~Consortium}, {Actis}, {Agnetta},
  {Aharonian}, {Akhperjanian} et~al.}]{2011ExA....32..193A}
\bibinfo{author}{{CTA~Consortium}}, \bibinfo{author}{M.~{Actis}},
  \bibinfo{author}{G.~{Agnetta}}, \bibinfo{author}{F.~{Aharonian}},
  \bibinfo{author}{A.~{Akhperjanian}}, et~al.,
\newblock \bibinfo{title}{{Design concepts for the Cherenkov Telescope Array
  CTA: an advanced facility for ground-based high-energy gamma-ray astronomy}},
\newblock \bibinfo{journal}{Experimental Astronomy} \bibinfo{volume}{32}
  (\bibinfo{year}{2011}) \bibinfo{pages}{193}.
%Type = Article
\bibitem[{Rowell et~al.(2008)Rowell, Stamatescu, Clay
  et~al.}]{2008NIMPA.588...48R}
\bibinfo{author}{G.~Rowell}, \bibinfo{author}{V.~Stamatescu},
  \bibinfo{author}{R.~Clay}, et~al.,
\newblock \bibinfo{title}{Tenten: A new \mbox{IACT} array for multi-{T}e{V}
  gamma-ray astronomy},
\newblock \bibinfo{journal}{Nuclear Instruments and Methods in Physics Research
  A} \bibinfo{volume}{588} (\bibinfo{year}{2008}) \bibinfo{pages}{48}.
%Type = Article
\bibitem[{{Gabici} and {Aharonian}(2007)}]{2007ApJ...665L.131G}
\bibinfo{author}{S.~{Gabici}}, \bibinfo{author}{F.~A. {Aharonian}},
\newblock \bibinfo{title}{{Searching for Galactic Cosmic-Ray Pevatrons with
  Multi-{T}e{V} Gamma Rays and Neutrinos}},
\newblock \bibinfo{journal}{The Astrophysical Journal Letters}
  \bibinfo{volume}{665} (\bibinfo{year}{2007}) \bibinfo{pages}{L131}.
%Type = Article
\bibitem[{{Tluczykont} et~al.(2011){Tluczykont}, {Hampf}, {Horns}
  et~al.}]{2011AdSpR..48.1935T}
\bibinfo{author}{M.~{Tluczykont}}, \bibinfo{author}{D.~{Hampf}},
  \bibinfo{author}{D.~{Horns}}, et~al.,
\newblock \bibinfo{title}{{The ground-based large-area wide-angle
  {$\gamma$}-ray and cosmic-ray experiment {HiSCORE}}},
\newblock \bibinfo{journal}{Advances in Space Research} \bibinfo{volume}{48}
  (\bibinfo{year}{2011}) \bibinfo{pages}{1935}.
%Type = Article
\bibitem[{{Ritt} et~al.(2010){Ritt}, {Dinapoli}, and
  {Hartmann}}]{2010NIMPA.623..486R}
\bibinfo{author}{S.~{Ritt}}, \bibinfo{author}{R.~{Dinapoli}},
  \bibinfo{author}{U.~{Hartmann}},
\newblock \bibinfo{title}{{Application of the DRS chip for fast waveform
  digitizing}},
\newblock \bibinfo{journal}{Nuclear Instruments and Methods in Physics Research
  A} \bibinfo{volume}{623} (\bibinfo{year}{2010}) \bibinfo{pages}{486}.
%Type = Inproceedings
\bibitem[{Serrano et~al.(2009)Serrano, Alvarez, Cattin, Cota, Lewis, Wlostowski
  et~al.}]{white_rabbit}
\bibinfo{author}{J.~Serrano}, \bibinfo{author}{P.~Alvarez},
  \bibinfo{author}{M.~Cattin}, \bibinfo{author}{E.~G. Cota},
  \bibinfo{author}{P.~M.~J.~H. Lewis}, \bibinfo{author}{T.~Wlostowski}, et~al.,
\newblock \bibinfo{title}{The white rabbit project},
\newblock in: \bibinfo{booktitle}{Proceedings of ICALEPCS TUC004, Kobe, Japan}.
  \bibinfo{note}{Also see: http://www.ohwr.org/projects/white-rabbit/}.
%Type = Inproceedings
\bibitem[{M.Brueckner and R.Wischnewski(2013)}]{brueckner_wr}
\bibinfo{author}{M.Brueckner}, \bibinfo{author}{R.Wischnewski},
\newblock \bibinfo{title}{A white rabbit setup for sub-nsec synchronization,
  timestamping and time calibration in large scale astroparticle physics
  experiments},
\newblock in: \bibinfo{booktitle}{Proceedings of the 33rd ICRC, Rio de
  Janeiro}. \bibinfo{note}{To appear, paper 1146}.
%Type = Inproceedings
\bibitem[{Brueckner et~al.(2013)Brueckner, Wischnewski, Berezhnev
  et~al.}]{brueckner_wr_2}
\bibinfo{author}{M.~Brueckner}, \bibinfo{author}{R.~Wischnewski},
  \bibinfo{author}{S.~Berezhnev}, et~al.,
\newblock \bibinfo{title}{Results from the whiterabbit sub-nsec time
  synchronization setup at {HiSCORE}-{T}unka},
\newblock in: \bibinfo{booktitle}{Proceedings of the 33rd ICRC, Rio de
  Janeiro}. \bibinfo{note}{To appear, paper 1158}.
%Type = Misc
\bibitem[{Wischnewski et~al.(2013)Wischnewski, Berezhnev, Brueckner
  et~al.}]{wischnewski_wr}
\bibinfo{author}{R.~Wischnewski}, \bibinfo{author}{S.~Berezhnev},
  \bibinfo{author}{M.~Brueckner}, et~al., \bibinfo{title}{Status of the
  {HiSCORE} experiment}, \bibinfo{year}{2013}. \bibinfo{note}{To appear, paper
  1164}.
%Type = Article
\bibitem[{{Budnev} et~al.(2008){Budnev}, {Chvalaev}, {Gress}
  et~al.}]{2008apsp.conf..287B}
\bibinfo{author}{N.~M. {Budnev}}, \bibinfo{author}{O.~B. {Chvalaev}},
  \bibinfo{author}{O.~A. {Gress}}, et~al.,
\newblock \bibinfo{title}{{Data Acquisition System for the TUNKA-133 Array}},
\newblock \bibinfo{journal}{Astroparticle, Particle and Space Physics,
  Detectors and Medical Physics Applications}  (\bibinfo{year}{2008})
  \bibinfo{pages}{287}.
%Type = Article
\bibitem[{{Schr{\"o}der} et~al.(2010){Schr{\"o}der}, {Asch}, {B{\"a}hren}
  et~al.}]{2010NIMPA.615..277S}
\bibinfo{author}{F.~G. {Schr{\"o}der}}, \bibinfo{author}{T.~{Asch}},
  \bibinfo{author}{L.~{B{\"a}hren}}, et~al.,
\newblock \bibinfo{title}{{New method for the time calibration of an
  interferometric radio antenna array}},
\newblock \bibinfo{journal}{Nuclear Instruments and Methods in Physics Research
  A} \bibinfo{volume}{615} (\bibinfo{year}{2010}) \bibinfo{pages}{277}.
%Type = Misc
\bibitem[{{Heck} et~al.(2012){Heck}, {Pierog}, and
  {Knapp}}]{2012ascl.soft02006H}
\bibinfo{author}{D.~{Heck}}, \bibinfo{author}{T.~{Pierog}},
  \bibinfo{author}{J.~{Knapp}}, \bibinfo{title}{{CORSIKA: An Air Shower
  Simulation Program}}, \bibinfo{year}{2012}. \bibinfo{note}{Astrophysics
  Source Code Library}.
%Type = Article
\bibitem[{{Kalmykov} et~al.(1997){Kalmykov}, {Ostapchenko}, and
  {Pavlov}}]{1997NuPhS..52...17K}
\bibinfo{author}{N.~N. {Kalmykov}}, \bibinfo{author}{S.~S. {Ostapchenko}},
  \bibinfo{author}{A.~I. {Pavlov}},
\newblock \bibinfo{title}{{Quark-Gluon-String Model and EAS Simulation Problems
  at Ultra-High Energies}},
\newblock \bibinfo{journal}{Nuclear Physics B Proceedings Supplements}
  \bibinfo{volume}{52} (\bibinfo{year}{1997}) \bibinfo{pages}{17}.
%Type = Misc
\bibitem[{Fesefeldt(1985)}]{fesefeldt:1985a}
\bibinfo{author}{H.~Fesefeldt}, \bibinfo{year}{1985}. \bibinfo{note}{Report
  PITHA-85/02, RWTH Aachen}.
%Type = Article
\bibitem[{Pierog(2013)}]{Pierog:2013qdx}
\bibinfo{author}{T.~Pierog},
\newblock \bibinfo{title}{{LHC data and extensive air showers}},
\newblock \bibinfo{journal}{European Physical Journal Web of Conferences}
  \bibinfo{volume}{52} (\bibinfo{year}{2013}) \bibinfo{pages}{03001}.
%Type = Misc
\bibitem[{{Hampf} et~al.(2010){Hampf}, {Tluczykont}, and
  {Horns}}]{2010tsra.confE.245H}
\bibinfo{author}{D.~{Hampf}}, \bibinfo{author}{M.~{Tluczykont}},
  \bibinfo{author}{D.~{Horns}}, \bibinfo{title}{{Simulation of the expected
  performance for the proposed gamma-ray detector {HiSCORE}}},
  \bibinfo{year}{2010}. \bibinfo{note}{Proceedings of Science (Texas 2010)
  245}.
%Type = Article
\bibitem[{{Bernl{\"o}hr}(2008)}]{2008APh....30..149B}
\bibinfo{author}{K.~{Bernl{\"o}hr}},
\newblock \bibinfo{title}{{Simulation of imaging atmospheric {C}herenkov
  telescopes with CORSIKA and sim\_telarray}},
\newblock \bibinfo{journal}{Astroparticle Physics} \bibinfo{volume}{30}
  (\bibinfo{year}{2008}) \bibinfo{pages}{149}.
%Type = Misc
\bibitem[{Hampf(2012)}]{hampf:phd}
\bibinfo{author}{D.~Hampf}, \bibinfo{title}{Study for the wide-angle air
  cherenkov detector {HiSCORE} and time gradient event reconstruction for the
  {H.E.S.S.} experiment}, \bibinfo{year}{2012}. \bibinfo{note}{PhD thesis,
  University of Hamburg, http://ediss.sub.uni-hamburg.de/volltexte/2012/5699/}.
%Type = Misc
\bibitem[{Kneizys et~al.(1996)Kneizys, Abreu, Anderson et~al.}]{modtran}
\bibinfo{author}{F.~Kneizys}, \bibinfo{author}{L.~Abreu},
  \bibinfo{author}{G.~Anderson}, et~al., \bibinfo{title}{The modtran 2/3 report
  and lowtran 7 model}, \bibinfo{year}{1996}. \bibinfo{note}{Edited by: Abreu,
  L.W., Anderson, G.P.}
%Type = Misc
\bibitem[{Henke(1994)}]{henke:dip}
\bibinfo{author}{V.~Henke}, \bibinfo{title}{Studie zur {A}uswertung der
  {A}nkunftszeitverteilung des {C}erenkov-{L}ichts ausgedehnter
  {L}uftschauer.}, \bibinfo{year}{1994}. \bibinfo{note}{Diploma thesis,
  University of Hamburg}.
%Type = Article
\bibitem[{{H{\"o}randel}(2003)}]{2003APh....19..193H}
\bibinfo{author}{J.~R. {H{\"o}randel}},
\newblock \bibinfo{title}{{On the knee in the energy spectrum of cosmic rays}},
\newblock \bibinfo{journal}{Astroparticle Physics} \bibinfo{volume}{19}
  (\bibinfo{year}{2003}) \bibinfo{pages}{193}.
%Type = Article
\bibitem[{{Hampf} et~al.(2011){Hampf}, {Rowell}, {Wild}, {Sudholz}, {Horns},
  and {Tluczykont}}]{2011AdSpR..48.1017H}
\bibinfo{author}{D.~{Hampf}}, \bibinfo{author}{G.~{Rowell}},
  \bibinfo{author}{N.~{Wild}}, \bibinfo{author}{T.~{Sudholz}},
  \bibinfo{author}{D.~{Horns}}, \bibinfo{author}{M.~{Tluczykont}},
\newblock \bibinfo{title}{{Measurement of night sky brightness in southern
  Australia}},
\newblock \bibinfo{journal}{Advances in Space Research} \bibinfo{volume}{48}
  (\bibinfo{year}{2011}) \bibinfo{pages}{1017}.
%Type = Article
\bibitem[{Nakamura et~al.(2010)}]{2010PDG}
\bibinfo{author}{K.~Nakamura}, et~al.,
\newblock \bibinfo{title}{Particle {D}ata {G}roup},
\newblock \bibinfo{journal}{Journal of Physics G} \bibinfo{volume}{37}
  (\bibinfo{year}{2010}) \bibinfo{pages}{075021}.
%Type = Article
\bibitem[{{Hampf} et~al.(2013){Hampf}, {Tluczykont}, and
  {Horns}}]{2013NIMPA.712..137H}
\bibinfo{author}{D.~{Hampf}}, \bibinfo{author}{M.~{Tluczykont}},
  \bibinfo{author}{D.~{Horns}},
\newblock \bibinfo{title}{{Event reconstruction techniques for the wide-angle
  air Cherenkov detector {HiSCORE}}},
\newblock \bibinfo{journal}{Nuclear Instruments and Methods in Physics Research
  A} \bibinfo{volume}{712} (\bibinfo{year}{2013}) \bibinfo{pages}{137}.
%Type = Article
\bibitem[{{Li} and {Ma}(1983)}]{1983ApJ...272..317L}
\bibinfo{author}{T.-P. {Li}}, \bibinfo{author}{Y.-Q. {Ma}},
\newblock \bibinfo{title}{{Analysis methods for results in gamma-ray
  astronomy}},
\newblock \bibinfo{journal}{The Astrophysical Journal} \bibinfo{volume}{272}
  (\bibinfo{year}{1983}) \bibinfo{pages}{317}.
%Type = Article
\bibitem[{Aharonian et~al.(2006)Aharonian, Akhperjanian, Aye
  et~al.}]{2006ApJ...636..777A}
\bibinfo{author}{F.~Aharonian}, \bibinfo{author}{A.~Akhperjanian},
  \bibinfo{author}{K.-M. Aye}, et~al.,
\newblock \bibinfo{title}{The {H.E.S.S.} survey of the inner galaxy in very
  high-energy gamma-rays},
\newblock \bibinfo{journal}{The Astrophysical Journal} \bibinfo{volume}{636}
  (\bibinfo{year}{2006}) \bibinfo{pages}{777}.
%Type = Article
\bibitem[{{Bartoli} et~al.(2013){Bartoli}, {Bernardini}, {Bi}
  et~al.}]{2013ApJ...779...27B}
\bibinfo{author}{B.~{Bartoli}}, \bibinfo{author}{P.~{Bernardini}},
  \bibinfo{author}{X.~J. {Bi}}, et~al.,
\newblock \bibinfo{title}{{TeV Gamma-Ray Survey of the Northern Sky Using the
  ARGO-YBJ Detector}},
\newblock \bibinfo{journal}{The Astrophysical Journal} \bibinfo{volume}{779}
  (\bibinfo{year}{2013}) \bibinfo{pages}{27}.
%Type = Article
\bibitem[{{Abdo} et~al.(2007){Abdo}, {Allen}, {Berley}
  et~al.}]{2007ApJ...658L..33A}
\bibinfo{author}{A.~A. {Abdo}}, \bibinfo{author}{B.~{Allen}},
  \bibinfo{author}{D.~{Berley}}, et~al.,
\newblock \bibinfo{title}{{Discovery of {T}e{V} Gamma-Ray Emission from the
  Cygnus Region of the Galaxy}},
\newblock \bibinfo{journal}{The Astrophysical Journal} \bibinfo{volume}{658}
  (\bibinfo{year}{2007}) \bibinfo{pages}{L33}.
%Type = Article
\bibitem[{{Meyer} et~al.(2010){Meyer}, {Horns}, and
  {Zechlin}}]{2010A&A...523A...2M}
\bibinfo{author}{M.~{Meyer}}, \bibinfo{author}{D.~{Horns}},
  \bibinfo{author}{H.-S. {Zechlin}},
\newblock \bibinfo{title}{{The Crab Nebula as a standard candle in very
  high-energy astrophysics}},
\newblock \bibinfo{journal}{Astronomy and Astrophysics} \bibinfo{volume}{523}
  (\bibinfo{year}{2010}) \bibinfo{pages}{A2}.
%Type = Article
\bibitem[{{Abdo} et~al.(2007){Abdo}, {Allen}, {Berley}
  et~al.}]{2007ApJ...664L..91A}
\bibinfo{author}{A.~A. {Abdo}}, \bibinfo{author}{B.~{Allen}},
  \bibinfo{author}{D.~{Berley}}, et~al.,
\newblock \bibinfo{title}{{{T}e{V} Gamma-Ray Sources from a Survey of the
  Galactic Plane with Milagro}},
\newblock \bibinfo{journal}{The Astrophysical Journal} \bibinfo{volume}{664}
  (\bibinfo{year}{2007}) \bibinfo{pages}{L91}.
%Type = Article
\bibitem[{{Aharonian} et~al.(2009){Aharonian}, {Akhperjanian}, {Anton}
  et~al.}]{2009A&A...499..723A}
\bibinfo{author}{F.~{Aharonian}}, \bibinfo{author}{A.~G. {Akhperjanian}},
  \bibinfo{author}{G.~{Anton}}, et~al.,
\newblock \bibinfo{title}{{Detection of very high energy radiation from HESS
  J1908+063 confirms the Milagro unidentified source MGRO J1908+06}},
\newblock \bibinfo{journal}{Astronomy and Astrophysics} \bibinfo{volume}{499}
  (\bibinfo{year}{2009}) \bibinfo{pages}{723}.
%Type = Article
\bibitem[{{Bartoli} et~al.(2012){Bartoli}, {Bernardini}, {Bi}
  et~al.}]{2012ApJ...760..110B}
\bibinfo{author}{B.~{Bartoli}}, \bibinfo{author}{P.~{Bernardini}},
  \bibinfo{author}{X.~J. {Bi}}, et~al.,
\newblock \bibinfo{title}{{Observation of the TeV Gamma-Ray Source MGRO
  J1908+06 with ARGO-YBJ}},
\newblock \bibinfo{journal}{The Astrophysical Journal} \bibinfo{volume}{760}
  (\bibinfo{year}{2012}) \bibinfo{pages}{110}.
%Type = Article
\bibitem[{{Tluczykont} et~al.(2012){Tluczykont}, {Horns}, {Hampf}, {Nachtigall}
  et~al.}]{2012NIMPA.692..246T}
\bibinfo{author}{M.~{Tluczykont}}, \bibinfo{author}{D.~{Horns}},
  \bibinfo{author}{D.~{Hampf}}, \bibinfo{author}{R.~{Nachtigall}}, et~al.,
\newblock \bibinfo{title}{{{HiSCORE}: A new detector for astroparticle and
  particle physics beyond 10 {T}e{V}}},
\newblock \bibinfo{journal}{Nuclear Instruments and Methods in Physics Research
  A} \bibinfo{volume}{692} (\bibinfo{year}{2012}) \bibinfo{pages}{246}.
%Type = Article
\bibitem[{Dubus et~al.(2013)Dubus, Contreras, Funk
  et~al.}]{2013APh....43..317D}
\bibinfo{author}{G.~Dubus}, \bibinfo{author}{J.~Contreras},
  \bibinfo{author}{S.~Funk}, et~al.,
\newblock \bibinfo{title}{Surveys with the {C}herenkov {T}elescope {A}rray},
\newblock \bibinfo{journal}{Astroparticle Physics} \bibinfo{volume}{43}
  (\bibinfo{year}{2013}) \bibinfo{pages}{317}.
%Type = Inproceedings
\bibitem[{{Sinnis}(2005)}]{2005AIPC..745..234S}
\bibinfo{author}{G.~{Sinnis}},
\newblock \bibinfo{title}{{HAWC: A Next Generation VHE All-Sky Telescope}},
\newblock in: \bibinfo{editor}{F.~A. {Aharonian}}, \bibinfo{editor}{H.~J.
  {V{\"o}lk}}, \bibinfo{editor}{D.~{Horns}} (Eds.),
  \bibinfo{booktitle}{Heidelberg Gamma-Ray Symposium}, volume
  \bibinfo{volume}{745} of \textit{\bibinfo{series}{American Institute of
  Physics}}, p. \bibinfo{pages}{234}.
%Type = Article
\bibitem[{{Cao}(2010)}]{2010ChPhC..34..249C}
\bibinfo{author}{Z.~{Cao}},
\newblock \bibinfo{title}{{A future project at {T}ibet: the large high altitude
  air shower observatory ({LHAASO})}},
\newblock \bibinfo{journal}{Chinese Physics C} \bibinfo{volume}{34}
  (\bibinfo{year}{2010}) \bibinfo{pages}{249}.
%Type = Article
\bibitem[{Cui et~al.(2014)Cui, Liu, Liu, and Ma}]{Cui201486}
\bibinfo{author}{S.~Cui}, \bibinfo{author}{Y.~Liu}, \bibinfo{author}{Y.~Liu},
  \bibinfo{author}{X.~Ma},
\newblock \bibinfo{title}{Simulation on gamma ray astronomy research with
  {LHAASO}-{KM2A}},
\newblock \bibinfo{journal}{Astroparticle Physics} \bibinfo{volume}{54}
  (\bibinfo{year}{2014}) \bibinfo{pages}{86 -- 92}.
%Type = Article
\bibitem[{{Gonzalez-Garcia} et~al.(2009){Gonzalez-Garcia}, {Halzen}, and
  {Mohapatra}}]{2009APh....31..437G}
\bibinfo{author}{M.~C. {Gonzalez-Garcia}}, \bibinfo{author}{F.~{Halzen}},
  \bibinfo{author}{S.~{Mohapatra}},
\newblock \bibinfo{title}{{Identifying Galactic PeVatrons with neutrinos}},
\newblock \bibinfo{journal}{Astroparticle Physics} \bibinfo{volume}{31}
  (\bibinfo{year}{2009}) \bibinfo{pages}{437}.

\end{thebibliography}

\end{document}